\documentclass[conference]{IEEEtran}
\usepackage[ruled,lined,linesnumbered,vlined]{algorithm2e}

\SetCommentSty{mycommfont}
\DontPrintSemicolon % Do not show the semicolons.

\SetKwProg{Fn}{Function}{:}{}

\usepackage{amsmath}
\usepackage{amsfonts}
\usepackage{amsthm}
\AtBeginDocument{%
  \setlength\abovedisplayskip{3pt plus 2pt minus 1pt}%
  \setlength\belowdisplayskip{3pt plus 2pt minus 1pt}%
  \setlength\abovedisplayshortskip{0pt plus 2pt}%
  \setlength\belowdisplayshortskip{3pt plus 2pt minus 1pt}%
}
\usepackage{balance}
\usepackage{booktabs}
\usepackage{enumitem}
\usepackage{graphicx}
\usepackage{listings}
\usepackage{multicol}
\usepackage{multirow}
\usepackage{subcaption}
\usepackage{url}
\usepackage{xspace}
\usepackage{xcolor}
\usepackage{tcolorbox}
\usepackage{colortbl}

\definecolor{lst-keyword}{HTML}{0000AF}
\definecolor{lst-string}{HTML}{A31515}
\definecolor{lst-comment}{HTML}{3A6B00}

\usepackage{tikz}
\usepackage{pgfplots}

\usetikzlibrary{shapes}
\usetikzlibrary{shapes.geometric}
\usetikzlibrary{arrows.meta, positioning}

\newcommand{\eg}{\mbox{\textit{e.g.}}\xspace}

\newcommand{\etc}{\mbox{\textit{etc.}}\xspace}
\newcommand{\ie}{\mbox{\textit{i.e.}}\xspace}
\definecolor{BlindColorTolOne}{HTML}{332288}
\definecolor{BlindColorTolTwo}{HTML}{117733} % green
\definecolor{BlindColorTolThree}{HTML}{44AA99}
\definecolor{BlindColorTolFour}{HTML}{88CCEE}
\definecolor{BlindColorTolFive}{HTML}{DDCC77}
\definecolor{BlindColorTolSix}{HTML}{CC6677} % light red
\definecolor{BlindColorTolSeven}{HTML}{AA4499}
\definecolor{BlindColorTolEight}{HTML}{882255}

\definecolor{BlindColorWongOne}{HTML}{000000} % black
\definecolor{BlindColorWongTwo}{HTML}{E69F00}
\definecolor{BlindColorWongThree}{HTML}{56B4E9}
\definecolor{BlindColorWongFour}{HTML}{009E73}
\definecolor{BlindColorWongFive}{HTML}{F0E442}
\definecolor{BlindColorWongSix}{HTML}{0072B2} % blue
\definecolor{BlindColorWongSeven}{HTML}{D55E00}
\definecolor{BlindColorWongEight}{HTML}{CC79A7}
\definecolor{mygreen}{HTML}{02818a}

\mathchardef\mhyphen="2D

\newcounter{FindingCounter}
\newcommand{\finding}[1]{%
	\begin{tcolorbox}[boxsep=1pt,top=2pt,bottom=2pt,left=4pt,right=4pt,
		before skip=3pt,after skip=3pt]
		\textbf{RQ\refstepcounter{FindingCounter}\theFindingCounter}: #1
	\end{tcolorbox}%
}

\newcommand{\myparagraph}[1]{
  \noindent \textit{\textbf{#1.}}\quad
}

\newcommand{\mycode}[1]{\lstinline{#1}}

\SetKwData{AlgTrue}{true}
\SetKwData{AlgFalse}{false}
\SetKw{AlgBreak}{break}
\SetKw{AlgContinue}{continue}

\newcounter{myUniqueIdCounter}

\makeatletter
\newcommand{\myGetOrAssignID}[1]{%
  \ifcsname myMap@#1\endcsname%
    \csname myMap@#1\endcsname%
  \else%
    \stepcounter{myUniqueIdCounter}%
    \expandafter\xdef\csname myMap@#1\endcsname{\themyUniqueIdCounter}%
    \themyUniqueIdCounter%
  \fi%
}
\makeatother

\newcommand{\anonymizedId}[1]{\ifx\useAnonymizedId\undefined%
  #1%
\else%
  \myGetOrAssignID{#1}%
\fi}

\newcommand{\dataviz}{DataViz\xspace}
\newcommand{\matplotlib}{matplotlib\xspace}
\newcommand{\bokeh}{bokeh\xspace}
\newcommand{\seaborn}{seaborn\xspace}
\newcommand{\ggplot}{ggplot2\xspace}
\newcommand{\vegaLite}{Vega-Lite\xspace}
\newcommand{\plotly}{plotly\xspace}

\newcommand{\Script}{\ensuremath{S}\xspace}
\newcommand{\Image}{\ensuremath{I}\xspace}
\newcommand{\RenderTree}{\ensuremath{T}\xspace}

\newcommand{\newbugs}{\ensuremath{47}\xspace}
\newcommand{\newbugsmpl}{\ensuremath{20}\xspace}
\newcommand{\newbugsbokeh}{\ensuremath{18}\xspace}
\newcommand{\newbugsplotly}{\ensuremath{9}\xspace}
\newcommand{\confirmedbugs}{\ensuremath{39}\xspace}
\newcommand{\confirmedbugsmpl}{\ensuremath{20}\xspace}
\newcommand{\confirmedbugsbokeh}{\ensuremath{13}\xspace}
\newcommand{\confirmedbugsplotly}{\ensuremath{6}\xspace}
\newcommand{\confirmedbugsvisual}{\ensuremath{34}\xspace}
\newcommand{\confirmedbugscrash}{\ensuremath{5}\xspace}
\newcommand{\confirmedbugslayout}{\ensuremath{11}\xspace}
\newcommand{\confirmedbugsannotaiton}{\ensuremath{12}\xspace}
\newcommand{\confirmedbugsscale}{\ensuremath{5}\xspace}
\newcommand{\confirmedbugsencoding}{\ensuremath{11}\xspace}
\newcommand{\pendingbugs}{\ensuremath{14}\xspace}
\newcommand{\pendingbugsmpl}{\ensuremath{8}\xspace}
\newcommand{\pendingbugsbokeh}{\ensuremath{4}\xspace}
\newcommand{\pendingbugsplotly}{\ensuremath{2}\xspace}

\newcommand{\fixedbugs}{\ensuremath{18}\xspace}
\newcommand{\fixedbugsmpl}{\ensuremath{11}\xspace}
\newcommand{\fixedbugsbokeh}{\ensuremath{7}\xspace}
\newcommand{\fixedbugsplotly}{\ensuremath{0}\xspace}

\newcommand{\toolname}{\textsc{VizDetour}\xspace}

\let\oldmycode\mycode
\renewcommand{\mycode}[1]{\ifmmode\text{\oldmycode{#1}}\else\oldmycode{#1}\fi}
\providecommand{\Node}{\ensuremath{\mathit{Node}}}
\providecommand{\Children}{\ensuremath{\mathit{Children}}}
\providecommand{\Properties}{\ensuremath{\mathit{Props}}}
\providecommand{\Prop}{\ensuremath{\mathit{prop}}}
\providecommand{\Elem}{\ensuremath{\mathit{elem}}}
\providecommand{\Script}{\ensuremath{\mathcal{S}}}

\providecommand{\NullDist}{\ensuremath{\mathit{NullDist}}}

\providecommand{\cpu}{AMD Ryzen Threadripper 3970X\xspace}
\providecommand{\mem}{256\,GiB\xspace}
\providecommand{\gpu}{NVIDIA RTX 6000 Ada Generation\xspace}
\providecommand{\os}{AlmaLinux 10.1\xspace}
\providecommand{\kernel}{6.12.0\xspace}

\providecommand{\pyver}{Python 3.13.11\xspace}
\providecommand{\mplver}{3.10.8\xspace}
\providecommand{\bokehver}{3.10.0\xspace}
\providecommand{\plotlyver}{6.8.0\xspace}
\providecommand{\numseedmpl}{\ensuremath{934}\xspace}
\providecommand{\numseedbokeh}{\ensuremath{471}\xspace}
\providecommand{\numseedplotly}{\ensuremath{1{,}082}\xspace}
\providecommand{\numseeds}{\ensuremath{2{,}487}\xspace}
\providecommand{\budgetK}{\ensuremath{10}\xspace}
\providecommand{\nullN}{\ensuremath{38{,}158}\xspace}
\providecommand{\timebudget}{\ensuremath{24}\xspace}

\providecommand{\baselineA}{Atheris\xspace}
\providecommand{\baselineB}{Fuzz4All\xspace}
\usepackage{cleveref} % This package must be loaded in the end.

\Crefname{algocf}{Algorithm}{Algorithms}
\crefname{algocf}{Algorithm}{Algorithms}

\Crefname{algorithm}{Algorithm}{Algorithms}
\crefname{algorithm}{Algorithm}{Algorithms}

\crefname{appendix}{Appendix}{Appendices}
\Crefname{appendix}{Appendix}{Appendices}

\Crefname{figure}{Figure}{Figures}
\crefname{figure}{Fig.}{Figs.}

\crefname{listing}{Listing}{Listings}
\Crefname{listing}{Listing}{Listings}

\Crefname{table}{Table}{Tables}
\crefname{table}{Table}{Tables}

\crefname{thm}{Theorem}{Theorems}
\Crefname{thm}{Theorem}{Theorems}

\crefname{equation}{Equation}{Equations}
\Crefname{equation}{Equation}{Equations}

\crefformat{chapter}{\S~#2#1#3}
\crefmultiformat{chapter}{\S\S~#2#1#3}{ and~#2#1#3}{, #2#1#3}{, and~#2#1#3}

\crefformat{section}{\S~#2#1#3}
\crefmultiformat{section}{\S\S~#2#1#3}{ and~#2#1#3}{, #2#1#3}{, and~#2#1#3}

\usepackage{cite}
\usepackage{amsmath,amssymb,amsfonts}
\usepackage{graphicx}
\usepackage{textcomp}
\usepackage{xcolor}
\usepackage{lipsum}
\usepackage{forest}

\def\BibTeX{{\rm B\kern-.05em{\sc i\kern-.025em b}\kern-.08em
    T\kern-.1667em\lower.7ex\hbox{E}\kern-.125emX}}
\begin{document}

%\title{Conference Paper Title*\\
%{\footnotesize \textsuperscript{*}Note: Sub-titles are not captured for https://ieeexplore.ieee.org  and
\title{Detecting Rendering Bugs in Imperative Data Visualization Libraries via Equivalent Mutations}

%\author{\IEEEauthorblockN{Anonymous Authors}}
\author{\IEEEauthorblockN{Weiqi Lu}
\IEEEauthorblockA{
\textit{Hong Kong University of Science and Technology}\\
Hong Kong, China \\
wluak@connect.ust.hk}
\and
\IEEEauthorblockN{Yongqiang Tian}
\IEEEauthorblockA{
\textit{Monash University}\\
Melbourne, Australia \\
yongqiang.tian@monash.edu}
%\and
%\IEEEauthorblockN{Shing-Chi Cheung}
%\IEEEauthorblockA{
%\textit{Hong Kong University of Science and Technology}\\
%Hong Kong, China \\
%scc@cse.ust.hk}
%\and
%\IEEEauthorblockN{Yuanmin Xie}
%\IEEEauthorblockA{
%\textit{Hong Kong University of Science and Technology}\\
%Hong Kong, China \\
%xieym3@gmail.com}
%\and
%\IEEEauthorblockN{Chengnian Sun}
%\IEEEauthorblockA{
%	\textit{University of Waterloo}\\
%	Waterloo, Canada \\
%	cnsun@uwaterloo.ca}
}

\maketitle

\begin{abstract}
Imperative data visualization libraries construct plots through a sequence of stateful API calls that incrementally create and update graphic elements. Rendering bugs in these libraries often manifest as incorrect visual outputs rather than crashes or exceptions, making them difficult to detect automatically. A fundamental challenge is the lack of an oracle that specifies the expected rendering of an arbitrary plotting script. Furthermore, an update to one graphic element may inadvertently affect other elements or properties, leading to subtle inconsistencies in the final rendered image.
%\scc{I cannot follow "unrelated property elsewhere in the internal state". I tried to rewrite. Pls check.} \wq{ok.}

This paper presents \toolname, an automated testing approach for detecting rendering bugs in imperative data visualization libraries via equivalent mutations. The key idea is to transform the oracle problem into an equivalence-checking problem. Starting from a seed plotting script, \toolname appends a short sequence of semantically equivalent API calls that temporarily modify the visualization state and then restore it to its original state. Although these mutations exercise different execution paths, they should preserve the final rendering. Any visual discrepancy between the original and mutated scripts therefore indicates a rendering bug. To generate such mutations, \toolname constructs a render tree from the seed script, identifies traceable graphic elements and mutable properties, and synthesizes endpoint-preserving mutation sequences. It then compares the rendered outputs using perceptual hashing. We evaluate \toolname on \matplotlib, \bokeh, and \plotly using scripts collected from their official example galleries. \toolname discovers \newbugs previously unknown bugs, of which \confirmedbugs are confirmed and \fixedbugs are fixed.
\end{abstract}

\begin{IEEEkeywords}
automatic software testing, data visualization, mutation-based testing, visual bug detection.
\end{IEEEkeywords}

\section{Introduction}

Data visualization (\dataviz) libraries are a fundamental component of
modern data analysis pipelines. They turn raw data into
graphical figures, or \emph{plots}, that make
patterns, comparisons, and structures visually accessible.
Libraries such as \matplotlib~\cite{tosi2009matplotlib},
\bokeh~\cite{jolly2018hands}, \plotly~\cite{Kruchten_An_interactive_open-source_2026}, \seaborn~\cite{waskom2021seaborn},
\vegaLite~\cite{satyanarayan2016vega}, and \ggplot~\cite{wickham2011ggplot2}
are widely used in statistical analysis~\cite{enache2023data}, exploratory data science~\cite{li2023edassistant},
academic publication~\cite{zadeh2024text2chart31}, dashboards~\cite{jiang2022mod2dash}, and GUI-based applications~\cite{bala2024python}, where
plots serve as the primary interface between raw data and human
interpretation.
Internally, these libraries expose APIs that transform data into visual encodings (\eg, heights, positions, colors) and encapsulate them into graphic elements (\eg, lines, bars, points). These elements are further annotated with labels, legends, and axes, forming a hierarchy.
Users then read these plots by combining perceptual understanding of
the visual encodings with textual understanding of the annotations,
mapping them back to the underlying data to recognize
distributions, trends, and anomalies.
Since users rely on such visualizations as evidence for interpreting
data, the correctness of
\dataviz libraries is critical: a silently mis-rendered plot can
lead users to misinterpret the underlying data, draw incorrect
conclusions, or make faulty decisions.
%\yq{I think we can diretly start wtih data visaluzation library, instead of the data visulaization.} \wq{addressed.}

Implementation defects in \dataviz libraries are particularly
insidious. Unlike crashes or exceptions, which surface immediately,
visualization bugs typically produce a plausible-looking but
semantically incorrect plot. A user may write a correct plotting
script, invoke the intended APIs, and inspect the resulting plots
assuming the library faithfully realizes the
requested encodings. If the library erroneously mutates visual properties, mishandles interactions between APIs, or wrongly propagates updates through stateful graphic elements, the resulting plots can still look correct while incorrectly capturing user intent. Recent empirical evidence~\cite{lu2025empirical} confirms that
such silent errors are widespread in mainstream \dataviz libraries,
span multiple components (\eg, visual encodings~\cite{githubHexbinBroken}, annotations~\cite{githubBugAdditive}, layouts~\cite{githubBugconstrained_layout}, scales~\cite{githubBugAxhist}), and frequently escape the
regression test suites. The prevalence and stealth of these defects
motivate systematic testing of \dataviz libraries. Designing an
effective methodology, however, requires understanding how
these libraries are structured and the defect patterns they exhibit.
%\yq{we should cite our emprical paper properly here to support our claim} \wq{cited.}
%\yq{question: the second source is the most common source or not? I think we do not need to mention the first one actually.} \wq{removed.}

\dataviz libraries fall into two categories that differ
substantially in programming model.
\emph{Imperative} libraries (\eg, \matplotlib, \bokeh) expose stateful
APIs that incrementally construct and mutate graphic elements.
\emph{Declarative} libraries (\eg, \seaborn, \vegaLite, \ggplot)
specify a high-level mapping from data to visual representations that
is subsequently evaluated or compiled into rendering operations.
We focus on imperative libraries for three reasons.
First, they serve as rendering backends for much of the \dataviz
ecosystem~\cite{towardsdatascienceDeclarativeImperative}, so their defects propagate to downstream declarative libraries, making them high-impact testing targets.
Second, imperative and declarative scripts follow different API patterns: imperative scripts rely on stateful calls that mutate graphic elements, while declarative scripts follow a grammar of graphics~\cite{wilkinson2011grammar}.
A single testing methodology is unlikely to address both effectively.
Third, the dominant bug patterns diverge accordingly: imperative libraries are uniquely prone to errors in state management, particularly initial element construction and subsequent update propagation~\cite{lu2025empirical}. Conversely, declarative libraries are prone to errors in specification translation and semantic mapping. Concentrating on imperative \dataviz libraries thus enables a focused and internally consistent methodology and evaluation.

\myparagraph{Motivating defect patterns}
As imperative \dataviz libraries build plots through sequences of stateful API calls, each call incrementally updates the internal element hierarchy. Locally sensible updates are therefore prone to violating non-local invariants among graphic elements. As a result, incorrect updating of visual properties is the second most common root cause of bugs in  \dataviz libraries~\cite{lu2025empirical}.
We examined 74 bugs from prior work~\cite{lu2025empirical} and found 44 (59.5\%) reproducible by appending API calls that update properties or modify the element hierarchy. The prevalence of such defects, exposed through subsequent state mutations, motivates our work.
%\yq{I feel part of the last two paragraph are actually the motivation of our study.
%	and I feel is kind of blur the focus of this background section. Do you think we can have a motivation subsection somewhere, and move some concepts here to that section? }
%\wq{I moved the last two paragraphs to the introduction section.}

\begin{figure*}[h]
	\centering
	% ---------- Column 1: Code ----------
	\begin{minipage}[c]{0.533\textwidth}
\begin{lstlisting}[
		language=python,
		breaklines=true,
		escapeinside={(*@}{@*)},
		emph={MarkerStyle,Affine2D,plot,get_fillstyle,set_fillstyle},
		emphstyle=\color{BlindColorWongSix}\bfseries,
		]
fig, ax = plt.subplots(figsize=(8, 8))
markers = []
for i in range(grid_size):
	for j in range(grid_size):
		x, y = x_coords[i, j], y_coords[i, j]
		angle = wind_directions[i, j]
		msize = 10 + wind_speeds[i, j] * 1.5
		marker = MarkerStyle('$\\rightarrow$', transform=Affine2D().rotate_deg(angle))
		line, = ax.plot(x, y, marker=marker, markersize=msize, color='#2c3e50', linestyle='None')
		markers.append((line, humidity_levels[i, j], line.get_fillstyle()))
# (*@\textbf{\textcolor{BlindColorWongSeven}{Appended update}}@*): encode humidity in fillstyle
for line, humidity, _ in markers:
	line.set_fillstyle('full' if humidity == 'high' else 'none')
# (*@\textbf{\textcolor{BlindColorWongSeven}{Appended restore}}@*): return fillstyle to its original
for line, _, original_fillstyle in markers:
	line.set_fillstyle(original_fillstyle)
\end{lstlisting}
		\centering{\footnotesize Bug-reproducing script (data generation omitted).}
	\end{minipage}\hfill
	% ---------- Column 2: Expected behavior (a) -> (b) -> (c) ----------
	\begin{minipage}[c]{0.233\textwidth}
		\centering
		\begin{subfigure}{\linewidth}
			\centering
			% mark this image so we can draw the equivalence arrow to (d)
			\tikz[remember picture,baseline]{\node[anchor=base,inner sep=0pt] (figA) {\includegraphics[width=0.68\linewidth]{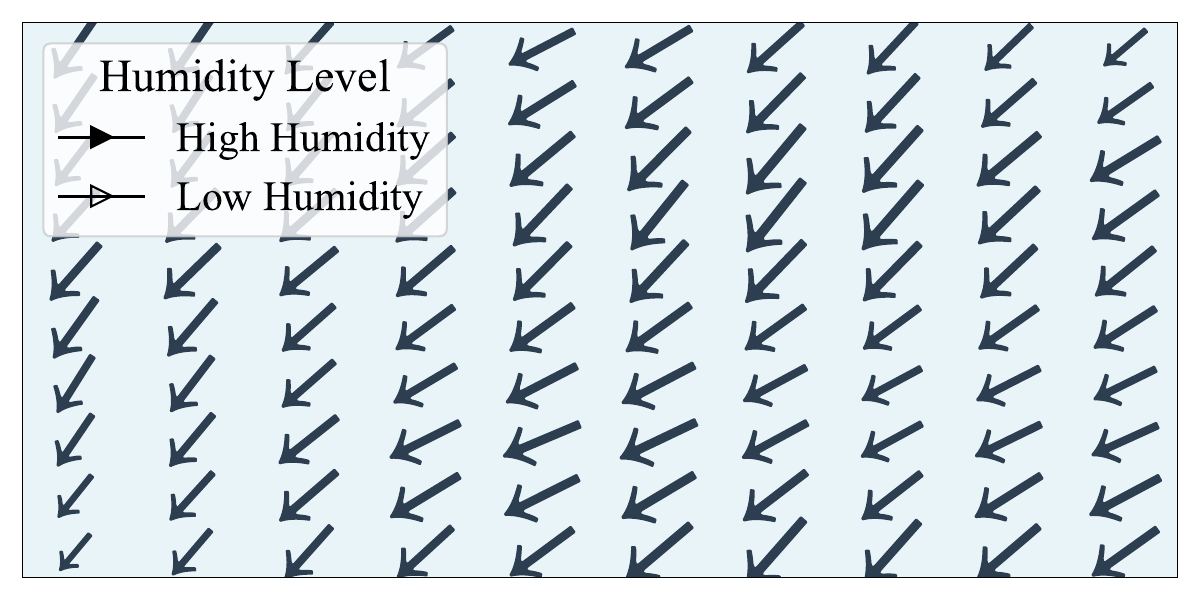}};}
			\subcaption{Expected original plot.}
			\label{subfig:motivating-original}
		\end{subfigure}\\[0.3em]
		$\downarrow$ \footnotesize Update Fillstyles \\[0.3em]
		\begin{subfigure}{\linewidth}
			\centering
			\includegraphics[width=0.68\linewidth]{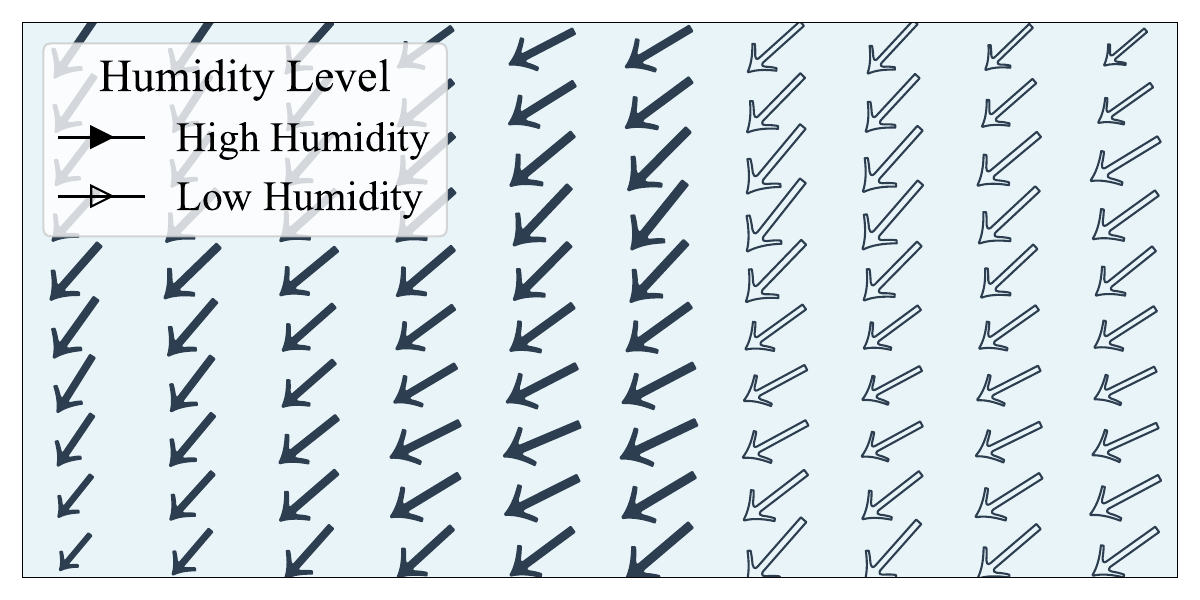}
			\subcaption{Expected updated plot.}
			\label{subfig:motivating-expected}
		\end{subfigure}\\[0.3em]
		$\downarrow$ \footnotesize Restore Fillstyles \\[0.3em]
		\begin{subfigure}{\linewidth}
			\centering
			\includegraphics[width=0.68\linewidth]{images/motivating-original.pdf}
			\subcaption{Expected restored plot.}
			\label{subfig:motivating-expected-revert}
		\end{subfigure}
	\end{minipage}\hfill
	% ---------- Column 3: Actual behavior (d) -> (e) -> (f) ----------
	\begin{minipage}[c]{0.233\textwidth}
		\centering
		\begin{subfigure}{\linewidth}
			\centering
			% mark this image as the target of the equivalence arrow from (a)
			\tikz[remember picture,baseline]{\node[anchor=base,inner sep=0pt] (figD) {\includegraphics[width=0.68\linewidth]{images/motivating-original.pdf}};}
			\subcaption{Actual original plot.}
			\label{subfig:motivating-original-2}
		\end{subfigure}\\[0.3em]
		$\downarrow$ \footnotesize Update Fillstyles \\[0.3em]
		\begin{subfigure}{\linewidth}
			\centering
			\includegraphics[width=0.68\linewidth]{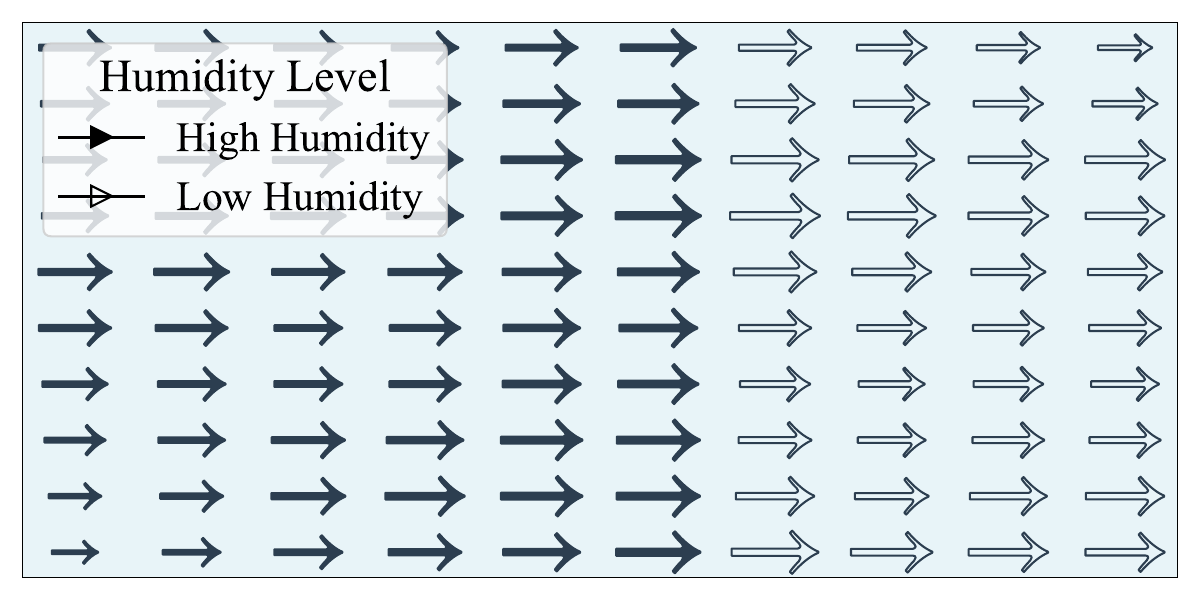}
			\subcaption{Actual updated plot.}
			\label{subfig:motivating-buggy}
		\end{subfigure}\\[0.3em]
		$\downarrow$ \footnotesize Restore Fillstyles \\[0.3em]
		\begin{subfigure}{\linewidth}
			\centering
			\includegraphics[width=0.68\linewidth]{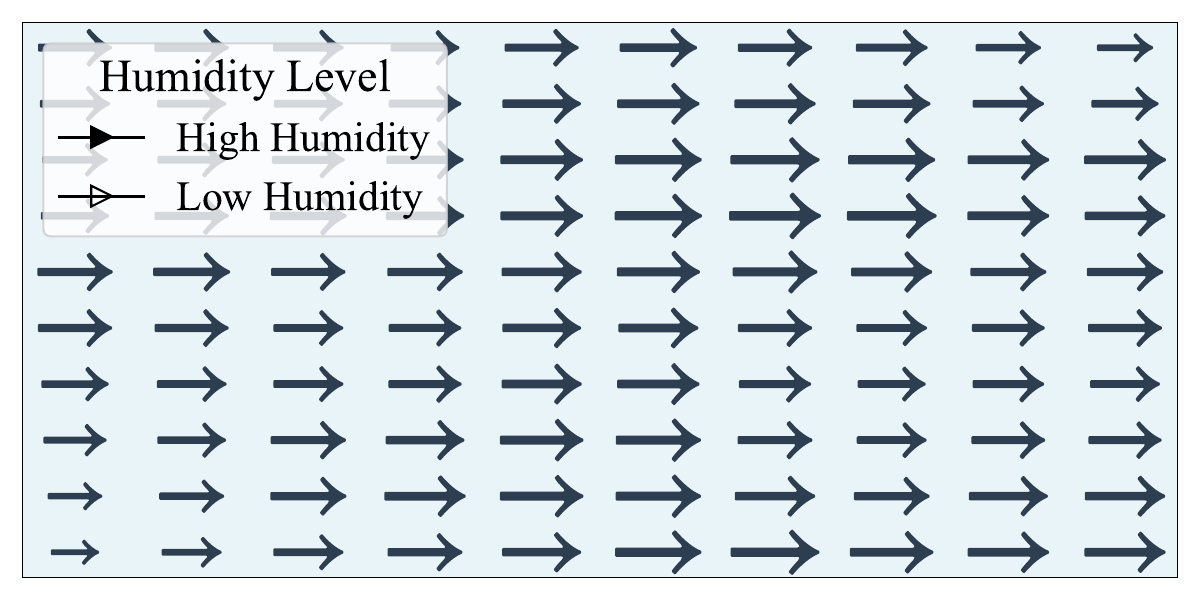}
			\subcaption{Actual restored plot.}
			\label{subfig:motivating-buggy-revert}
		\end{subfigure}
	\end{minipage}
	% ---------- Equivalence arrow between (a) and (d) ----------
	\begin{tikzpicture}[remember picture,overlay]
		% horizontal connector between the two images
		\draw[<->,thick,shorten >=3pt,shorten <=3pt]
		(figA.east) -- (figD.west);
		% "equiv." label in a white box so it masks the line
		\node[font=\footnotesize,fill=white,inner sep=0pt,yshift=4pt]
		at ($(figA.east)!0.5!(figD.west)$) {equiv.};
	\end{tikzpicture}
	\caption{
		Motivating example adapted from a \matplotlib bug~\cite{githubBugUpdate} detected by \toolname:
		updating marker fillstyles unexpectedly resets the rotation
		transforms of markers. The left column shows the plotting script.
		Both transitions start from the \emph{same} original input
		(\cref{subfig:motivating-original}~$\equiv$~\cref{subfig:motivating-original-2}).
		The middle column shows the
		\emph{expected} transitions: updating fillstyles only restyles the
		arrows~(\cref{subfig:motivating-expected}), and the restored plot matches the
		original~(\cref{subfig:motivating-expected-revert}~$\equiv$~\cref{subfig:motivating-original}).
		The right column shows the \emph{actual} buggy behavior: updating
		fillstyles silently strips the rotation of every marker~(\cref{subfig:motivating-buggy}), and the corruption persists
		after restoring~(\cref{subfig:motivating-buggy-revert}~$\not\equiv$~\cref{subfig:motivating-original-2}).
		While checking only the updated fillstyle misses the hidden rotation corruption, a straightforward equality comparison between the original and restored plots successfully exposes the bug.
		%		\yq{should we say that figure a and b are the same? in b's caption, we can say original input as (a)}
		%		\wq{I added an equivalent line between (a) and (d)}
		%		\yq{too tiny.}
		%		\wq{updated.}
		%		\yq{need to update}
		%		\wq{updated.}
		%		\wq{Todo: Remove title. Reduce height. Add fig.c, fig.a.... Arrange vertically, merge code.}
		%		\wq{Updated.}
		%		\wq{Todo: Add fig.c, fig.a.... in the remaining paragraphs.}
		%		\wq{Finished.}
	}
	\label{fig:motivating-example}
\end{figure*}

\myparagraph{A motivating example}
To illustrate how stateful update defects manifest,
the left column of \cref{fig:motivating-example} presents a bug-reproducing script
adapted from \matplotlib Issue~\#31257~\cite{githubBugUpdate}, uncovered by
\toolname.
%\yq{if it s a bug found by us, then we should mention this.} \wq{mentioned.}
The script visualizes a
synthetic wind field on a regular grid. At each grid cell, an arrow
marker (\mycode{'\$\\\\rightarrow\$'}) is rotated by the local wind
direction via an \mycode{Affine2D} transform, sized by wind speed, and
added to the plot through \mycode{ax.plot}. A subsequent loop encodes humidity
into the \textit{fillstyle} of each marker by calling \mycode{set\_fillstyle('full')} for
high-humidity cells and \mycode{set\_fillstyle('none')} otherwise.
After this loop, the user expects arrows to keep their local wind-direction rotation
while fillstyle encodes humidity~(\cref{subfig:motivating-expected}).
Instead, \matplotlib silently removes marker rotation, snapping arrows to the rightward
orientation~(\cref{subfig:motivating-buggy}). All API calls are locally correct, no warning
is raised, and the plot still looks plausible.

The defect is rooted in the stateful update logic for marker styles.
Each \mycode{line} stores its marker as a \mycode{MarkerStyle} instance, which encapsulates the marker shape, the \mycode{Affine2D} transform,
and the fillstyle. When \mycode{set_fillstyle(fs)} is invoked, the
library reconstructs a fresh \mycode{MarkerStyle} instead of updating
the existing instance in place. Because this reinitialization starts
from the marker shape alone, the previously attached rotation transform
is silently dropped.

\myparagraph{Challenges}
Detecting the motivating defect, and silent visualization defects of
this kind more generally, through automated testing poses three
challenges.

\emph{C1: Demand for a generalized visual oracle.} For an arbitrary plotting
script, there is no general specification of what the rendered image
\emph{should} look like, so absolute image-level oracles are
unavailable~\cite{lu2025empirical,6963470}.
Manually authored reference images cover only a tiny
fraction of the input space and cannot generalize to mutated or
synthesized scripts.

\emph{C2: Requirement for syntactically and semantically valid API sequences.} Triggering the bug requires a specific sequence of multiple interacting API calls, not a single one. Furthermore, this sequence must mirror real-world data visualization practices. Random fuzzing with unreasonable or undocumented API combinations may generate corrupted plots rather than exposing genuine bugs~\cite{csmith, DBLP:journals/software/BohmeCR21}. Even when a random fuzzer happens to follow the documentation and trigger a valid defect, the unconstrained mutation of unrelated properties and elements can obscure the buggy component. For instance, a mutation might render a large rectangle over it.

\emph{C3: Necessity for validating both related and unrelated properties.} Validating a stateful API update must confirm both that the target property changes and that unrelated properties remain unchanged. In our example, after \mycode{set_fillstyle}, this means checking fillstyle and invariance of color, markersize, transform, linestyle, z-order, \etc Checking only fillstyle misses the bug because corruption appears in an unrelated rotation property. But exhaustive checks after every API call are impractical: the number of elements grows with plot complexity, and properties vary by library and version.

\myparagraph{Insight}
We observe that defects in stateful update logic can be exposed without
predicting each update's correct outcome. The key is to
extend the original execution with an additional update path that should
return to a state semantically equivalent to the
terminal state. The rendered plot after this path should therefore match
the original plot. Any residual corruption introduced along the path manifests
as a difference between the two images.
Taking~\cref{fig:motivating-example} as an example, we append code that
restores the fillstyle after the original update. The
restored plot should be identical to the original plot. This holds
for the correct execution
(\cref{subfig:motivating-expected-revert}~$\equiv$~\cref{subfig:motivating-original}).
However, the actual restored plot differs from the original plot because
the rotation loss persists
(\cref{subfig:motivating-buggy-revert}~$\not\equiv$~\cref{subfig:motivating-original-2}). This comparison exposes the defective state transition without an oracle
for the intermediate plot.

\myparagraph{Our approach}
We generalize this observation as \emph{endpoint-preserving mutation}. Given a seed script, \toolname appends a short sequence of
valid API calls that may traverse different intermediate states but should preserve the terminal state. Under this
construction, a correct library should render the seed script and the
mutated script identically. If the two rendered images differ,
\toolname reports the augmented script as a self-contained reproducing
script.

Concretely, \toolname approximates the internal state of a seed script
with its render-tree, a tree-structured representation of graphic
elements and their mutable properties. It samples traceable graphic
elements and properties to identify mutation targets. It then
synthesizes endpoint-preserving mutants via three operators:
\emph{set-revert}, \emph{redundant-set}, and \emph{remove-readd}. Each
mutant is executed, and its rendered image is compared against the seed
image by perceptual-hash distance.

This formulation resolves the three challenges as follows.
For~\emph{C1}, the endpoint-preserving construction replaces an
absolute oracle with a relative oracle: two executions that reach
semantically equivalent terminal states should produce
perceptually identical images. For~\emph{C2}, each mutation is appended to a real seed script
and operates only on elements and properties already present in the
plot. The augmented script is therefore a valid, usage-grounded API
sequence that exercises stateful update logic without obscuring the defect. For~\emph{C3}, endpoint preservation reduces
per-property validation to a single image comparison. Any corruption on a related or unrelated property surfaces as a
perceptual difference between the seed and restored plots.

\myparagraph{Contributions}
This paper contributes the following:
%\yq{I revised the contribution a litte. I feel the previous one is too long and there are some overlap.}
%\wq{ok}
\begin{itemize}[topsep=0pt, leftmargin=*]
	\item \textbf{Problem formulation.} We formalize imperative \dataviz library testing as a \emph{state-equivalence} problem, with endpoint-preserving mutation as the central test generation strategy.
	\item \textbf{Render-tree as state approximation.} We introduce the render-tree as a tractable approximation of \dataviz library internal state, and develop algorithms for render-tree construction and traceable element sampling.
	\item \textbf{Semantics-preserving mutation operators.} We design three semantics-preserving operators (\emph{set-revert}, \emph{redundant-set}, \emph{remove-readd}) to realize endpoint-preserving mutation.
	\item \textbf{Implementation and evaluation.} We implement the above as \toolname and evaluate it on \matplotlib, \bokeh, and \plotly. \toolname discovers \newbugs previously unknown bugs, of which \confirmedbugs are confirmed and \fixedbugs fixed. 
%	We make our implementation and detected-bug dataset publicly available~\cite{vizdetourRepo}.
%	\wq{todo: create a repo for data availability, include the testing framework, seeds, new bugs, evaluation results. https://github.com/smith2936/vizdetour}
\end{itemize}

\section{Preliminaries}
\label{sec:background}

%This section reviews the three concepts that form the foundation of this work:
%imperative \dataviz libraries and their stateful internal model
%(\Cref{sec:bg-imperative}), mutation-based test generation as a
%methodology for producing valid test inputs from existing scripts
%(\Cref{sec:bg-mutation}), and image-based test oracles for
%deciding whether two images are consistent (\Cref{sec:bg-oracle}).

This section introduce imperative \dataviz libraries,
mutation-based test generation, and image-based test oracles.
% as the foundation of this work.

\subsection{Imperative Data Visualization Libraries}
\label{sec:bg-imperative}

Imperative \dataviz libraries (\eg, \matplotlib, \bokeh) expose
stateful APIs through which scripts incrementally construct and configure a collection
of graphic elements (called \emph{artists} in \matplotlib~\cite{matplotlibArtists}, \emph{glyphs}
in \bokeh~\cite{bokehGlyphs}). A typical plotting script first allocates a figure and its axes,
then adds graphic elements via calls (\eg, \mycode{plot},
\mycode{add\_artist}, \mycode{add\_glyph}), and configures
their visual properties. Each call updates a hierarchy rooted at a
\mycode{Figure}.
The hierarchy's \emph{interior nodes} are composite elements (\eg, axes, legends,
containers) that group child elements and manage a shared configuration.
Its \emph{leaf nodes} are primitive elements (\eg, \mycode{Line2D}, \mycode{Patch},
\mycode{Point}) with concrete visual attributes such as geometry, color, transform,
and z-order~\cite{brown2012architecture}.
\textit{Declarative \dataviz libraries}
(\eg, \vegaLite, \ggplot), in contrast, compile an immutable
specification to pixels in one shot and expose no per-element
mutable state.

The stateful nature of imperative libraries provides a unique advantage for automated testing, exploiting two key architectural features.
%\yq{I am wondering shall we have this paragraph here or not? I feel it is too early to talk about this.} \wq{Ok. I rephrased getter-setter and add-remove as they should be discussed in the approach section.}
First, these libraries expose internal visual attributes via runtime mutable properties, allowing an external framework to both read and rewrite the state of any graphic element. Second, the structural hierarchy of the library's internal state is itself mutable, meaning that graphic elements can be added or removed after initial construction. Together, these property mutations and structural mutations define the state transitions that the library must keep internally consistent. These two mutable surfaces are the exact targets for the testing technique introduced in~\cref{sec:approach}.

%\yq{but here we have not introduce what is our approach formally. I feel hard to follow
%	the remaining sentences.}
%\wq{I avoided mentioning our approach in all the preliminaries.}

\subsection{Mutation-Based Test Generation}
\label{sec:bg-mutation}

Mutation-based test generation produces new inputs by transforming existing valid inputs~\cite{ye2023comfuzz, eom2024fuzzing, song2023metamong}.
It takes as input a corpus of \textit{seed} scripts that execute successfully and a set of \textit{mutation operator}s that rewrite a seed script by inserting, deleting, or replacing code fragments.
Each rewritten script is a \textit{mutant} that is subsequently executed and assessed by a test oracle~\cite{6963470}.
In automated testing, mutation reframes verification from absolute correctness to relative correctness~\cite{diallo2015correctness, alblwi2025subsumptionRelativeCorrectness},
requiring only that the mutant and seed satisfy the relation defined by the mutation operator.
%\wq{todo: add citations.}

%\yq{do we really need this paragraph? I feel no...} \wq{Updated}
For imperative \dataviz libraries, directly deleting or replacing fragments is often problematic because API calls and parameters are interdependent. Instead, we mimic deletion or replacement by inserting new statements rather than modifying existing ones. Since these changes can be appended to a completed plot without breaking prior dependencies, mutants are generated by inserting code fragments at the end of a seed script. The inserted calls are constructed to revert their intermediate effects, so the mutant remains equivalent to the seed script while exercising the library's stateful update logic.

\subsection{Image-Based Test Oracles}
\label{sec:bg-oracle}

Automated testing of visual systems requires an image-based oracle to decide whether rendered output matches expectations~\cite{alegroth2015visual, donaldson2017automated, mayer2006random}. Existing oracles range from pixel-level metrics such as MSE and SSIM~\cite{palubinskas2017image}, to image-processing methods~\cite{hassaballah2016image}, deep feature extractors~\cite{dara2018feature}, and vision-language models~\cite{roberts2024image2struct}. Although learning-based oracles capture high-level semantics, their cost, non-determinism, and runtime overhead make them poorly suited to high-throughput testing.

For mutation-based testing of imperative \dataviz libraries, the oracle must be efficient and robust to benign rendering noise~\cite{lu2025empirical}. Pixel-level comparisons are brittle because anti-aliasing, font hinting, and floating-point variation can trigger false positives between equivalent plots.
Perceptual hashing (pHash) offers a practical alternative by encoding an image as a compact binary fingerprint whose Hamming distance reflects perceptual similarity~\cite{zauner2010implementation}. Used in browser regression testing and GUI diffing~\cite{zhou2025janus, song2023metamong, yandrapally2020nearduplicate}, pHash tolerates minor visual noise while remaining sensitive to structural visualization faults such as missing elements, incorrect shapes, layouts, positions, colors, or corrupted plots. Thus, pHash provides a fast, deterministic oracle aligned with human perception and suitable for scalable \dataviz library testing.
%\yq{this reads like AI}
%\wq{Rewritten.}

\section{Approach}
\label{sec:approach}
%\input{figures/workflow.tex}
% workflow_v2.tex
\begin{figure*}[t]
	\centering
	\resizebox{0.9\textwidth}{!}{%
		\begin{tikzpicture}[
			font=\small,
			every node/.style={align=center},
			script/.style={
				draw, thick, fill=gray!8,
				minimum width=23mm, minimum height=10mm
			},
			run/.style={
				draw, thick, rounded corners=8pt,
				fill=blue!6,
				minimum width=16mm, minimum height=8mm
			},
			image/.style={
				draw, thick, fill=purple!6,
				minimum width=25mm, minimum height=12mm
			},
			module/.style={
				draw, thick, rounded corners=4pt,
				fill=blue!3,
				minimum width=85mm, minimum height=22mm
			},
			mutstep/.style={
				draw, thick, rounded corners=2pt,
				fill=white,
				minimum width=22mm, minimum height=10mm
			},
			badge/.style={
				draw, circle, thick,
				fill=blue!10,
				inner sep=1.4pt,
				font=\scriptsize\bfseries
			},
			oracle/.style={
				draw, thick, rounded corners=3pt,
				fill=orange!10,
				minimum width=28mm, minimum height=13mm
			},
			report/.style={
				draw, thick, rounded corners=8pt,
				fill=red!6,
				minimum width=30mm, minimum height=9mm
			},
			pass/.style={
				draw, thick, rounded corners=8pt,
				fill=green!10,
				minimum width=27mm, minimum height=9mm
			},
			flow/.style={-{Latex[length=2mm]}, thick},
			dashflow/.style={-{Latex[length=2mm]}, thick, dashed},
			note/.style={font=\footnotesize, inner sep=1pt}
			]
			%% ---------- Reference construction ----------
			\node[script] (seed) at (0,0)
			{Seed Script\\$\Script_0$};
			\node[run] (run0) at (2.4,0) {Run};
			\node[image] (img0) at (4.9,0)
			{Reference Image\\$\Image_0$};

			%% ---------- Mutation module ----------
			\node[module] (module) at (4.6,-2.35) {};
			\node[font=\bfseries] at (4.8,-1.64)
			{Render-tree guided endpoint-preserving mutation};
			\node[mutstep] (tree) at (1.8,-2.55)
			{Build\\render-tree\\$\RenderTree_{k-1}$};
			\node[badge] at (0.72,-2.05) {1};
			\node[mutstep] (sample) at (4.6,-2.55)
			{Sample\\$\langle \Elem,\Prop\rangle$};
			\node[badge] at (3.42,-2.05) {2};
			\node[mutstep] (mutate) at (7.4,-2.55)
			{Append mutation\\$\Delta_k$ to $\Delta_{\text{acc}}$};
			\node[badge] at (6.12,-2.05) {3};

			%% ---------- Follow-up execution ----------
			\node[script] (aug) at (10.45,-2.55)
			{Augmented Script\\$\Script_k = \Script_0\oplus\Delta_{\text{acc}}$};
			\node[run] (runk) at (13.1,-2.55) {Run};
			\node[image] (imgk) at (15.6,-2.55)
			{Follow-up Image\\$\Image_k$};

			%% ---------- Oracle and outcomes (Step 4) ----------
			\node[oracle] (oracle) at (15.6,0)
			{Visual Oracle\\$d(\Image_0,\Image_k)$};
			\node[badge] at (14.2,0.5) {4};   % <-- Step 4 badge added here

			\node[report] (report) at (18.9,0)
			{Report Incorrect Plot\\$d>\tau$};
			\node[pass] (pass) at (18.9,-1.15)
			{No Deviation\\$d\le\tau$};

			%% ---------- Edges ----------
			\draw[flow] (seed) -- (run0);
			\draw[flow] (run0) -- (img0);
			\draw[flow] (seed.south) |- (module.west);
			\draw[flow] (tree) -- (sample);
			\draw[flow] (sample) -- (mutate);
			\draw[flow] (mutate) -- (aug);
			\draw[flow] (aug) -- (runk);
			\draw[flow] (runk) -- (imgk);
			\draw[flow] (img0) -- (oracle);
			\draw[flow] (imgk.north) -- (oracle.south);
			\draw[flow] (oracle) -- (report);
			\draw[flow] (oracle.-25) -- (pass.west);

			%% ---------- Feedback Loop ----------
			\draw[dashflow] (pass.south) -- (18.9,-3.7) -- node[above, pos=0.25, font=\small] {Next iteration: $k \gets k+1$ (while $k < K$)} (1.8,-3.7) -- (tree.south);
	\end{tikzpicture}}
	\caption{Overview of \toolname. The framework executes a seed script
		\(\Script_0\) to obtain the reference image \(\Image_0\). It then
		uses the render-tree to create endpoint-preserving mutations that
		should keep the terminal state equivalent to the seed
		state. The augmented script produces a follow-up image
		\(\Image_k\), and the visual oracle compares
		\(\Image_k\) with \(\Image_0\). If the distance exceeds the
		calibrated threshold \(\tau\), \toolname reports an incorrect plot. Otherwise, the workflow
		continues until the budget \(K\) is exhausted.}
	\label{fig:overview-v2}
\end{figure*}
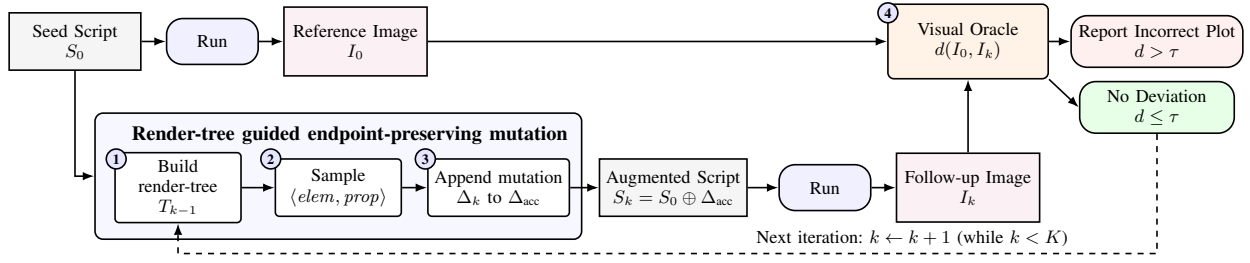

\begin{figure*}[h]
	\begin{subfigure}{0.45\textwidth}
\begin{lstlisting}[
		language=python,
		breaklines=true,
		escapeinside={(*@}{@*)},
		emph={MarkerStyle,Affine2D,plot,get_fillstyle,set_fillstyle},
		emphstyle=\color{BlindColorWongSix}\bfseries
		]

fig, ax = plt.subplots()
for x, theta in enumerate([-20, 0, 20]):
    marker = MarkerStyle('*', transform=Affine2D().rotate_deg(theta))
    ax.plot(x, 0, marker=marker, markersize=100)
\end{lstlisting}
		\subcaption{Seed script $\Script_0$ written in \matplotlib that generates the reference image $\Image_0$ in \cref{fig:running-example-expected}. Import statements are omitted}
		\label{fig:running-example-script}
	\end{subfigure}
	\begin{subfigure}{0.5\textwidth}
		\centering
		{\scriptsize
	\begin{forest}
		for tree={l sep=5pt, s sep=4pt, inner ysep=1pt}
		[Figure, text=red
		[Rectangle]
		[Axes, text=red, edge={red, thick}
		[{[Line2D]}, text=red, edge={red, thick}]
		[{[Spine]}]
		[XAxis]
		[YAxis]
		[{[Text]}]
		[Rectangle]
		]
		]
	\end{forest}
}
		\subcaption{Render-tree $\RenderTree_0$ of $\Script_0$. The red path marks the element sampled in \cref{fig:running-example-oracle-code}. Square brackets~([]) denote multiple elements of the same type, and nodes deeper than the third level are omitted for brevity.}
		\label{fig:running-example-tree}
	\end{subfigure}
	\begin{subfigure}{0.38\textwidth}
\begin{lstlisting}[
		language=python,
		breaklines=true,
		escapeinside={(*@}{@*)},
		emph={MarkerStyle,Affine2D,plot,get_fillstyle,set_fillstyle, get_children},
		emphstyle=\color{BlindColorWongSix}\bfseries
		]
# ... plotting script unchanged...
elem = fig.get_children()[1].get_children()[0]  # get the first Line2D marker (blue marker)
v = elem.get_fillstyle()
elem.set_fillstyle('left')
elem.set_fillstyle(v)
\end{lstlisting}
		\subcaption{Set-revert mutation $\Delta_1$ appended to $\Script_0$ to produce the augmented script $\Script_1$.}
		\label{fig:running-example-oracle-code}
	\end{subfigure}
	\hfill
	\begin{subfigure}{0.28\textwidth}
		\centering
		\includegraphics[width=\linewidth]{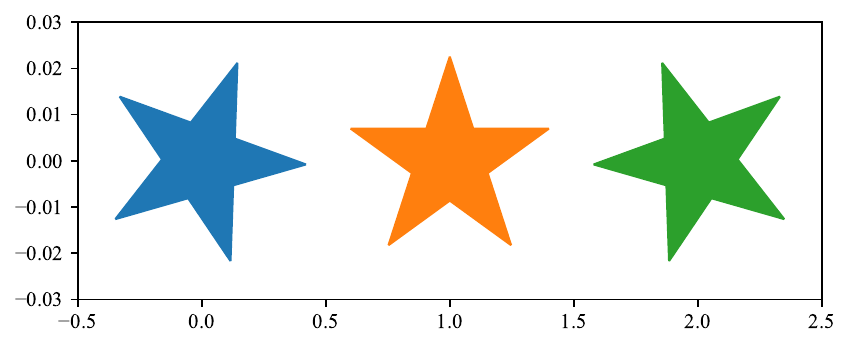}
		\subcaption{Reference image $I_0$ (expected output of the augmented script $\Script_1$).}
		\label{fig:running-example-expected}
	\end{subfigure}
	\hfill
	\begin{subfigure}{0.28\textwidth}
		\centering
		\begin{tikzpicture}
			% Render the original plot image
			\node[anchor=south west, inner sep=0] (image) at (0,0) {\includegraphics[width=\linewidth]{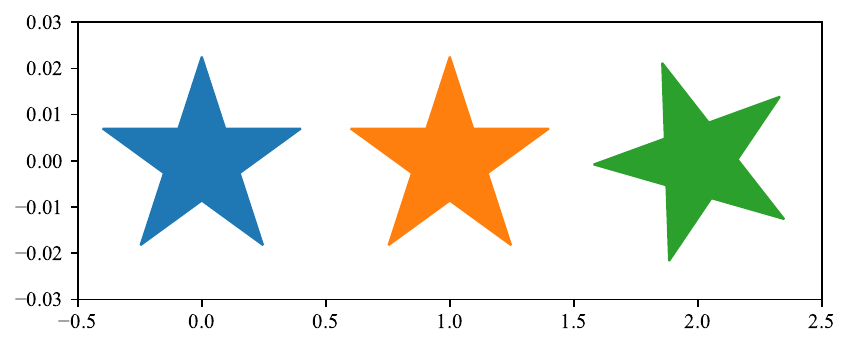}};
			\begin{scope}[x={(image.south east)},y={(image.north west)}]
				% Draw a red dashed ellipse over the first star (image is wide, so
				% use explicit radii to keep it round on the page)
				\draw[red, dashed, thick] (0.236, 0.542) ellipse [x radius=0.135, y radius=0.31];
				% Add a small text indicator pointing to it
				\node[red, font=\sffamily\tiny\bfseries, anchor=south] at (0.236, 0.87) {Rotation lost!};
			\end{scope}
		\end{tikzpicture}
		\subcaption{Actual output of the augmented script $\Script_1$ (rotation of the star is lost).}
		\label{fig:running-example-inconsistency}
	\end{subfigure}
	\caption{An Illustrative Example of the Testing Framework \toolname.}
	\label{fig:running-example}
\end{figure*}

This section presents \toolname, a framework for testing imperative \dataviz libraries. We formalize the testing problem via internal rendering states and endpoint-preserving mutations, then detail the core components of \toolname.

\cref{fig:overview-v2} shows the workflow of \toolname, and \cref{fig:running-example} illustrates it on a concrete \matplotlib script used as a running example throughout this section. Given a seed script $\Script_0$ (\eg, the script plotting three rotated star-shaped markers in \cref{fig:running-example-script}), \toolname proceeds in four stages.
In Step 1, it executes $\Script_0$ to obtain a reference image $\Image_0$ (\cref{fig:running-example-expected}) and builds a \emph{render-tree} $\RenderTree_0$ from the program's runtime state (detailed in \Cref{sec:rendertree}; see \cref{fig:running-example-tree}). The render-tree captures the hierarchy of graphic elements and provides programmatic access paths for later mutation.
In Step 2, \toolname uniformly samples a target graphic element $e\in\RenderTree_0$, one of its mutable properties, and an \emph{access expression} $p$ that can syntactically refer to $e$ from the seed script (detailed in \Cref{sec:sampling}). In the running example, it samples the first \mycode{Line2D} marker, addressable as \mycode{fig.get_children()[1].get_children()[0]}, and its \mycode{fillstyle} property.
In Step 3, \toolname synthesizes an \emph{endpoint-preserving mutation} $\Delta$ using the sampled element, property, and access expression $p$ (detailed in \Cref{sec:operators}), then appends it to the seed script. As shown in \cref{fig:running-example-oracle-code}, $\Delta$ reads the marker's current \mycode{fillstyle}, temporarily sets it to \mycode{'left'}, and immediately restores the original value, yielding an augmented script $\Script_k$ whose terminal state should match that of $\Script_0$.
In Step 4, \toolname executes $\Script_k$ to obtain a follow-up image $\Image_k$ and compares it with the reference image $\Image_0$ using a visual oracle based on image distance (detailed in \Cref{sec:oracle}). It repeats this process for up to $K$ rounds and reports $\Script_k$ as suspicious if the distance exceeds a threshold $\tau$. In the running example, one round ($k=1$) exposes the discrepancy: \matplotlib's \mycode{set_fillstyle} restores the marker's fill style but loses its \mycode{Affine2D} rotation, producing axis-aligned stars in $\Image_1$ instead of the rotated ones in $\Image_0$ (\cref{fig:running-example-inconsistency}). \Cref{sec:algo} presents the overall test generation algorithm that integrates these four stages.

\subsection{Problem Formalization}
\label{sec:formalization}

\myparagraph{Internal rendering state and its approximation}
An imperative \dataviz script \(\Script=\langle c_1,\ldots,c_n\rangle\) is a sequence of API calls that incrementally constructs a plot.
Each call updates the \emph{internal rendering state}, the program state the library uses to render the plot.
We distinguish two views of this state.
The \emph{expected} internal rendering state is the state that should hold according to the library documentation and the script.
The \emph{actual} internal rendering state is the concrete runtime program state.
Fully modeling the actual state is impractical due to its complex internal attributes and backend dependencies~\cite{devChoosingRight}.

We approximate the internal rendering state with a \emph{render-tree} that structurally captures the library's element hierarchy. A render-tree is a tree of graphic elements of a plot, rooted at the figure.
Its interior nodes are composite elements (\eg, axes, legends) that group child elements, and its leaf nodes are primitive elements (\eg, lines, markers).
Each node exposes the mutable visual properties of its element, and \Cref{sec:rendertree} details its construction.
The render-tree gives \toolname a structured view of these elements and the access expressions used to reference them from the script.
%\yq{Issue 1 (naming collision): the render-tree is introduced as our stand-in for the internal state, so a render-tree already \emph{is} a state. Yet ``render-tree state'' here names a different object, a group of render-trees. The word ``state'' thus denotes two things, which is why ``render-tree state'' reads oddly.} \wq{resolved. I removed ``render-tree state''.}
%\yq{Issue 2 (the definition makes a defect impossible): state-equivalence is defined by identical rendered \emph{output}. Under this definition two render-trees in the same state render identically by construction, so the core testing criterion below becomes a tautology and the running-example defect cannot occur. We likely want a semantic definition, \eg agreement on every mutable visual property, so that a defect is a state-equivalent pair that renders differently.} \wq{Revised.}

\myparagraph{State transitions and rendering}
Each call \(c_i\) induces a \emph{state transition} \(\sigma_{i-1}\xrightarrow{c_i}\sigma_i\). We distinguish the \emph{actual} transition the runtime library performs from the \emph{expected} transition the documentation and script prescribe. We refer to \(\pi(\Script)=\sigma_0\sigma_1\cdots\sigma_n\) as the \emph{execution path} of \(\Script\), where \(\sigma_0\) is the empty canvas, and the final state \(\sigma_n\) is the \emph{endpoint}. In \cref{fig:running-example-script}, the seed issues seven API calls, with the endpoint \(\sigma_7\).

Let $\sigma \equiv \sigma'$ denote equivalence between two internal rendering states. It holds if and only if corresponding graphic elements carry equal visual property values. We model rendering as a function \(R\colon\Sigma\to\mathcal{I}\) that maps a state to its rendered image, with state set \(\Sigma\) and image set \(\mathcal{I}\). A correctly implemented library must render equivalent states to perceptually identical images, \ie \(\sigma\equiv\sigma'\) implies \(R(\sigma)\approx R(\sigma')\). We use perceptual equality \(\approx\) rather than pixel equality because rendering admits benign noise (\Cref{sec:oracle}).

\myparagraph{From validating state to validating state change}
Deciding whether the actual endpoint or image matches its expected counterpart is hard, because both are difficult to derive from the documentation and script alone.
We instead validate a state change, which isolates the elements and properties targeted by a short update sequence. To keep validation tractable, we focus on state changes specified to return to an equivalent state, whose the rendered image should not change.

\myparagraph{Endpoint-preserving mutation}
Given a seed script $\Script_0$ with endpoint $\sigma_n$, \toolname synthesizes a sequence of additional calls $\Delta=\langle c_{n+1},\ldots,c_{n+m}\rangle$ that modify the sampled graphic elements.
%\yq{render-tree is the approxiation of the internal state, and now we leveraging it to modify the internal state. Emm, I feel not very right.} \wq{I removed render-tree as it is detailed in the approximation paragraph.}
We call $\Delta$ an \emph{endpoint-preserving mutation} when its calls are specified to restore the state, so that the \emph{expected} endpoint $\sigma_{n+m}^{\mathrm{exp}}$ of $\Script_0\oplus\Delta$ satisfies $\sigma_{n+m}^{\mathrm{exp}}\equiv\sigma_n$ by construction. The mutation $\Delta=\langle\mycode{set_fillstyle('left')},\mycode{set_fillstyle(v)}\rangle$ in \cref{fig:running-example-oracle-code} is endpoint-preserving, prescribing $\sigma_{n+2}^{\mathrm{exp}}\equiv\sigma_n$.
A correctly implemented
% \yq{correctly implemented} \wq{ok}
library reaches an \emph{actual} endpoint $\sigma_{n+m}^{\mathrm{act}}\equiv\sigma_n$, and hence renders $R(\sigma_{n+m}^{\mathrm{act}})\approx R(\sigma_n)$. Contrapositively, any $\Delta$ for which $R(\sigma_{n+m}^{\mathrm{act}})\not\approx R(\sigma_n)$ witnesses $\sigma_{n+m}^{\mathrm{act}}\not\equiv\sigma_n$.
%\yq{why it is $\approx$}
%\yq{have we defined R?}
Such a divergence exposes a defect in the library's stateful update logic. \toolname therefore uses image comparison as a visual oracle to surface these inconsistencies.

\subsection{Render-Tree Construction}
\label{sec:rendertree}

%\wq{introduce render-tree}
\toolname builds the render-tree from the runtime state of the library
after executing the seed script. It serves two purposes: enumerating candidate elements for
mutation and deriving an executable expression that refers to a
sampled element from the seed script.
Building the tree traverses the native object model of the specific library.
Imperative \dataviz libraries differ in how they expose this structure
(\eg, \matplotlib exposes an \mycode{Artist} hierarchy via
\mycode{get_children()}, while \bokeh exposes its component graph via
\mycode{children} and \mycode{renderers}).
%\yq{while xxx (another lib) xxxxxx} \wq{added \bokeh}
\toolname therefore confines this dependence to a single
\(\Children(e)\) hook, which returns the ordered immediate children of
\(e\) with the syntax to reference each child from the
script. Subsequent stages operate on this tree, making the rest of the framework library-agnostic.

\begin{algorithm}
	\footnotesize
	\caption{Render-Tree Construction}
	\label{alg:render_tree}
	\SetKwFunction{Build}{Build}
	\SetKwFunction{Children}{children}
	\KwIn{Root element \(r\), root access expression \(p_{root}\)}
	\KwOut{Render-tree root \(\RenderTree\)}
	\Fn{\Build{\(e, par, a\)}}{
		\(v \gets \textbf{new } \Node\)\;
		\(v.e \gets e,\quad v.C \gets [\ ],\quad v.par \gets par,\quad v.a \gets a\)\;

		\ForEach{\((e', a') \in \Children(e)\)}{
			\(u \gets \Build(e', v, a')\)\tcp*{\(a'\): accessor of child \(e'\)}
			\(v.C.\texttt{append}(u)\)\;
		}

		\Return \(v\)\;
	}
	\Return \Build(\(r, \textbf{null}, p_{root}\))\;
\end{algorithm}

\myparagraph{Tree representation}
The render-tree consists of nodes \(\Node(e,C,par,a)\), where
\(e\) is the underlying graphic element, \(C\) is the ordered list of
immediate child nodes, \(par\) is a pointer to the parent node, and
\(a\) is the accessor fragment that references \(e\) from its parent
(the root binding \(p_{root}\) for the root node).
We materialize this annotated tree rather than reuse the native
hierarchy, because the native object model exposes neither an
upward pointer to the parent nor the accessor syntax to reference
each element from the script. The sampler of \Cref{sec:sampling} relies
on the parent pointers and accessor fragments to recover the access
expression of any sampled node by a single walk to the root.
\Cref{alg:render_tree} builds this representation by a
depth-first traversal starting at the root \(r\) (\eg, the
\mycode{Figure} in \matplotlib), recording each parent pointer and
accessor fragment as it descends. Applied to the terminal state of
\cref{fig:running-example-script}, the procedure walks from the
root \mycode{Figure} to its children (the background \mycode{Rectangle}
and \mycode{Axes}), then recursively into the \mycode{Axes}, yielding
the tree in \cref{fig:running-example-tree} with
three \mycode{Line2D} markers alongside spine, axis, and
text elements.

\subsection{Element and Property Sampling}
\label{sec:sampling}

%\yq{Possible over-engineering: Algorithm~2 (the subtree-size-weighted descent) exists to sample a node uniformly in \(O(\depth)\) rather than \(O(N)\). But \(N\) is the number of elements in a plot (tens to low thousands), where an \(O(N)\) pass is instant, so the optimization targets a non-problem. The identical-distribution alternative: during one traversal, collect every element with its access expression into a flat list, then pick one uniformly. This would let us delete the subtree-size field \(n\) (used only by the sampler), collapse Algorithm~2 to a one-liner, and drop the \(O(\depth)\) complexity claim and the probability argument. Keep the render-tree concept (it earns its place in the formalization, and the root-to-node path \emph{is} the access expression), \(\Children\), and getter/setter property discovery. Caveat: after simplifying we should not claim any efficiency contribution, since uniform sampling over a small set is not one.} \wq{I changed it to sampling from a flatten list, and focus on element access expression generation based on the render-tree.}

Given the render-tree, this stage samples a single mutation target. Each target consists of three parts: the graphic element, its access expression in the script, and a mutable property.

\begin{algorithm}
	\footnotesize
	\caption{Uniform Node Sampling with Element Access Expression}
	\label{alg:uniform_sampling_elem_path}
	\SetKwFunction{Flatten}{Flatten}
	\SetKwFunction{AccessExpr}{AccessExpr}
	\SetKwFunction{Rand}{Rand}
	\KwIn{Render-tree root \(\RenderTree\)}
	\KwOut{Sampled node \(v\) and its element access expression \(p\)}
	\Fn{\Flatten{\(v, L\)}}{
		\(L.\texttt{append}(v)\)\;
		\lForEach{\(u \in v.C\)}{\Flatten{\(u, L\)}}
	}
	\Fn{\AccessExpr{\(v\)}}{
		\lIf{\(v.par = \textbf{null}\)}{\Return \(v.a\)\tcp*[f]{root binding \(p_{root}\)}}
		\Return \AccessExpr{\(v.par\)} \(+\ v.a\)\;
	}
	\(L \gets [\ ]\)\;
	\Flatten{\(\RenderTree, L\)}\tcp*{collect all nodes into a flat list}
	\(v \gets L[\Rand(0, |L|)]\)\tcp*{sample a node uniformly}
	\(p \gets \AccessExpr{v}\)\tcp*{recover access expression via parents}
	\Return \((v, p)\)\;
\end{algorithm}

\myparagraph{Uniform element sampling}
%\yq{Justification mismatch: uniform-over-nodes does NOT give different levels equal attention. Node counts per level are wildly unequal (1 figure, a few axes, but dozens of leaf ticks/spines/text/markers), so the sampling is dominated by the numerous leaves, and high-level containers are hit only with probability \(1/N\). The mechanism equalizes attention \emph{per element}, which is the opposite of \emph{per level}. Two options: (1) keep uniform-over-elements and delete the ``different levels receive equal attention'' claim (recommended, honest, still publishable); or (2) if per-level coverage is actually the goal, switch to a stratified scheme (sample a level/type first, then an element within it) and motivate why. Also note the joint distribution over (element, property) is uniform-over-elements then uniform-within, not jointly uniform.} \wq{I applied option 1.}
\Cref{alg:uniform_sampling_elem_path} samples every element uniformly, so each element in the render-tree, from high-level containers to individual primitives, has the same probability of selection. It flattens the render-tree into a list $L$ of all nodes by a depth-first traversal, then draws one node uniformly from $L$. Each node therefore has an exact $1/|L|$ probability of selection. Because building the render-tree requires traversing all elements, generating this flat list introduces negligible overhead.\toolname rebuilds the render-tree and resamples every round, because mutations may cause the \dataviz library to alter element hierarchy or ordering.

\myparagraph{Element access expression}
Beyond returning a node $v$, the sampler recovers an \emph{access expression} $p$ for it. When substituted directly into the seed script, this expression evaluates to $v.e$ within the script's existing execution flow. To construct the expression, the sampler walks upward from $v$ through the parent pointers to the root, concatenating the accessor fragment $a$ recorded at each node. This upward traversal prepends successive parent contexts to the accumulated accessor fragment, concluding at a library-specific root binding $p_{root}$ (\eg, \mycode{fig} in \matplotlib). In the running example, the sampled \mycode{Line2D} marker is resolved as \mycode{fig.get_children()[1].get_children()[0]} in \cref{fig:running-example-oracle-code}. This access expression ensures every generated mutation is a syntactically valid, self-contained extension of the seed script.

\myparagraph{Property sampling}
It remains to choose which property of the sampled element to perturb. For each element $e \in \RenderTree$, we collect its set of \emph{mutable properties} $\Properties(e)$ by matching getter-setter pairs through naming conventions (\eg, a \mycode{get_x} method with its \mycode{set_x} counterpart). For the \mycode{Line2D} element sampled in \cref{fig:running-example-oracle-code}, $\Properties(e)$ includes \mycode{fillstyle}, \mycode{markersize}, \mycode{color}, \mycode{linestyle}, \mycode{transform}, and thirty other properties. We draw one property uniformly at random from $\Properties(e)$ as the mutation target. For this property, we require only the weak contract that \mycode{set_x} accepts values of the same format that \mycode{get_x} returns. We do \emph{not} assume that \mycode{set_x(get_x())} is the identity operation on the underlying state. Violations of that stronger identity assumption are among the defects that the subsequent mutation aim to expose.

\subsection{Endpoint-Preserving Mutation Operators}
\label{sec:operators}

Given a sampled element \(e\) with access expression \(p\) and a
mutable property \(\Prop\in\Properties(e)\), \toolname instantiates one of
three mutation operators, each producing a \(\Delta\) that is
endpoint-preserving by construction.
%\yq{Operator-selection policy is unstated: we never say how one of the three operators is chosen. Need a sentence on applicability and selection, \eg RR applies only to elements with an \mycode{add}/\mycode{remove} pair, SR needs a property with candidate values, RS needs a getter--setter pair, and among the applicable operators one is chosen (uniformly at random? by priority?). Algorithm~3 just labels this step ``SR, RS, or RR'' without the policy.} \wq{Updated to random policy here and the \Cref{alg:main}.}
The mutation operators differ in applicability. Redundant-set operator applies to any property with a getter--setter pair, provided the setter accepts the same format that the getter returns. Set-revert additionally requires candidate values for the property type that are distinct from the original value.  Remove-readd applies only when the \mycode{add}/\mycode{remove} API pair is exposed by either the element itself or its parent element. Among the operators applicable to the sampled target, \toolname selects one uniformly at random.

\subsubsection{Set-Revert (SR)}
Let \(v=\mycode{get}_\Prop(p)\) be the current value of the property.
The set-revert operator emits
\[
\Delta_{\text{SR}}=\langle
\text{\color{blue}\textbf{\mycode{set}}}_{\Prop}(p,v'),\ \text{\color{blue}\textbf{\mycode{set}}}_{\Prop}(p,v)
\rangle,
\]
for some \(v'\neq v\) drawn from a set of candidate values selected based on the property type.
The second call restores the property to its
original value, so \(\sigma_{n+2}^{\mathrm{exp}}\equiv\sigma_n\) by construction.
\Cref{fig:running-example-oracle-code} is a set-revert instance
with \(\Prop=\mycode{fillstyle}\) and \(v'=\mycode{'left'}\). The
operator targets bugs where a library fails to re-derive equivalent internal rendering state after a property is modified and reset. In the running example, changing and reverting the fillstyle causes the library to reconstruct the marker from scratch, accidentally discarding its rotation and causing the visual defect in \cref{fig:running-example-inconsistency}.

\myparagraph{Mutation Value Selection}
The set-revert operator requires a candidate value \(v'\neq v\), drawn
from a type-dependent value pool. For a numeric property (\mycode{int},
\mycode{float}), we draw boundary values, special floats (\eg, infinity,
NaN), and scaled or offset perturbations of the original value. For a
boolean property, we set \(v'=\neg v\). For an enumeration property, we
parse the valid set from the library documentation or the runtime enum
definition and sample \(v'\neq v\) uniformly. For \mycode{fillstyle} in
\cref{fig:running-example-oracle-code}, \matplotlib documents
the valid set as \(\{\mycode{full},\mycode{left},\mycode{right},
\mycode{bottom},\mycode{top},\mycode{none}\}\), and since the current value
is \mycode{full}, the sampler drew \(v'=\mycode{left}\).
When the property type cannot be resolved through introspection or
documentation, we fall back to resampling \(v\) itself, which reduces
the operator to redundant-set and probes idempotency.

\subsubsection{Redundant-Set (RS)}
The redundant-set operator reapplies the value already held by the property:
\[
\Delta_{\text{RS}}=\langle \text{\color{blue}\textbf{\mycode{set}}}_{\Prop}(p, \text{\color{blue}\textbf{\mycode{get}}}_{\Prop}(p)) \rangle.
\]
Although trivially endpoint-preserving in specification, a correct implementation must be idempotent. We observe cases where such calls fail due to getter-setter incompatibilities~\cite{githubBugColorbar} or unintended side effects on other properties~\cite{githubBugAdjusting}.

\subsubsection{Remove-Readd (RR)}
%\yq{I made RR explicit about position restoration (capture original index \(j\), re-insert via \mycode{add}(e,j)) so that the endpoint-preserving claim actually holds: a plain \mycode{add}(e) would append at the end, changing draw order and the image even in a correct library, which would be a false positive. Please confirm the implementation really does this, and that the target libraries expose an \mycode{add} that accepts a position. If not, the formula and the \(\sigma^{\mathrm{exp}}\equiv\sigma_n\) argument need adjusting (or RR restricted to cases where order does not affect rendering).} \wq{RR restricted to cases where order does not affect rendering.}
For elements that can be manipulated via an \mycode{add}/\mycode{remove} API pair (\eg, \mycode{add_artist}/\mycode{remove}), the remove-readd operator evaluates the access expression $p$ to retrieve the target element reference \(e = \mycode{eval}(p)\), then emits
\[
\Delta_{\text{RR}}=\langle \text{\color{blue}\textbf{\mycode{remove}}}(e),\ \text{\color{blue}\textbf{\mycode{add}}}(e)\rangle.
\]
The \mycode{add} call re-inserts \(e\) at the end of the sibling order rather than at its original position, because the \mycode{add} APIs of the target libraries append by default. We therefore restrict this operator to elements whose draw order does not affect the rendered image. Under this restriction the re-added element carries the same visual property values and renders identically, so \(\sigma_{n+2}^{\mathrm{exp}}\equiv\sigma_n\) holds by construction and a correct library reproduces \(\Image_0\). Applying remove-readd to an element whose sibling order influences the rendered image would change the draw order and the image even in a correct library, so these cases are excluded.
This operator probes whether removing and re-adding an element correctly restores its original state and rendered image. It also doubles as a \emph{crash probe}: underlying bugs in internal state tracking, such as deleting a reference while other components still hold pointers to it, can cause \mycode{remove}(e) to crash the runtime, directly exposing a defect.

\subsection{Visual Oracle}
\label{sec:oracle}

The endpoint-preserving mutation reduces bug
detection to one decision: whether the seed and its
mutant render to perceptually identical images. The
visual oracle makes this decision. It takes the reference image
\(\Image_0\) rendered from the seed \(\Script_0\) and the follow-up image
\(\Image_k\) rendered from the augmented script
\(\Script_k=\Script_0\oplus\Delta\) (\Cref{sec:operators}). It then
operationalizes the relation \(\approx\) used to define perceptual identity.

Following \Cref{sec:bg-oracle}, we instantiate \(\approx\) with
perceptual hashing (pHash), which absorbs benign rendering noise while
remaining sensitive to the structural corruptions that characterize
stateful update defects. We compute the Hamming distance
\(d(\Image_0,\Image_k)\) between their perceptual hashes and report \(\Script_k\) as a
\emph{suspicious reproducing script} when \(d(\Image_0,\Image_k)>\tau\).
Because the appended mutation is endpoint-preserving by construction, a
correct library renders \(\Image_k\) perceptually identical to
\(\Image_0\) and satisfies \(d(\Image_0,\Image_k)\le\tau\). Any larger
distance is evidence of a defect in the library's stateful update
logic. The threshold \(\tau\)
is calibrated during evaluation (discussed in \Cref{subsec:oracle_cal}).
%\yq{one paragraph as a subsection?} \wq{improved.}

\subsection{Test Generation Algorithm}
\label{sec:algo}

\Cref{alg:main} assembles the components above into the full
test-generation loop. For each seed \(\Script_0\), we execute up to
\(K\) rounds of mutation, each targeting a freshly sampled element.
Each round rebuilds the render-tree~(\Cref{sec:rendertree}), then samples the mutation target in
two steps. \textsc{UniformNodeSampling} draws a node \(v\) uniformly
together with its access expression \(p\), and
\textsc{PropertySampling} draws one mutable property \(\Prop\) uniformly
from \(\Properties(v.e)\)~(\Cref{sec:sampling}). The round then selects an operator uniformly
at random from the subset of \{SR, RS, RR\} applicable to the sampled
element and property, and instantiates the
corresponding \(\Delta_k\)~(\Cref{sec:operators}).
Because the concatenation of endpoint-preserving mutations is itself
endpoint-preserving, the invariant \(\sigma_{\text{terminal}}^{\mathrm{exp}}\equiv\sigma_n\)
is maintained throughout the test. The loop halts early and emits a
reproducing script as soon as a suspicious rendered image is observed.
If no deviation is observed within \(K\) rounds, the seed passes under
the current budget. On \cref{fig:running-example-script}, the algorithm
terminates at \(k=1\) with the reproducing script of
\cref{fig:running-example-oracle-code}.
The accumulated \(\Delta_{\text{acc}}\), once appended to
\(\Script_0\), forms a self-contained, executable reproducing script.
Because every mutation is a syntactic extension of the seed script, it
runs against the library under test with no auxiliary harness,
letting developers reproduce the defect by executing it as is.

\begin{algorithm}[t]
	\footnotesize
	\caption{\textsc{GenerateTests}}
	\label{alg:main}
	\SetKwFunction{Execute}{Execute}
	\SetKwFunction{BuildTree}{BuildRenderTree}
	\SetKwFunction{NodeSample}{UniformNodeSampling}
	\SetKwFunction{PropSample}{PropertySampling}
	\SetKwFunction{Applicable}{ApplicableOps}
	\SetKwFunction{Rand}{Rand}
	\SetKwFunction{GenMutation}{GenMutation}
	\SetKwFunction{Report}{Report}
	\SetKwFunction{Pass}{Pass}
	\KwIn{Seed script \(\Script_0\), root access expression \(p_{root}\), budget \(K\), threshold \(\tau\)}
	\KwOut{A verdict with a reproducing script or \(\emptyset\)}
	\((r_0,\Image_0)\gets\Execute{\(\Script_0\)}\)\tcp*[l]{\(r_0\): root element, \(\Image_0\): reference image}
	\(\Delta_{\text{acc}}\gets\langle\rangle\)\;
	\For{\(k\gets 1\) \KwTo \(K\)}{
		\(\RenderTree_{k-1}\gets\BuildTree{\(r_{k-1}, p_{root}\)}\)\tcp*[l]{Alg.~\ref{alg:render_tree}}
		\((v,p)\gets\NodeSample{\(\RenderTree_{k-1}\)}\)\tcp*[l]{Alg.~\ref{alg:uniform_sampling_elem_path}}
		\(\Prop\gets\PropSample{\(v.e\)}\)\tcp*[l]{\Cref{sec:sampling}}
		\(op\gets\Rand(\Applicable{\(v.e,\Prop\)})\)\tcp*[l]{uniform over applicable \(\subseteq\)\,\{SR, RS, RR\}}
		\(\Delta_k\gets\GenMutation{\(v.e,p,\Prop,op\)}\)\;
		\(\Delta_{\text{acc}}\gets\Delta_{\text{acc}}\oplus\Delta_k\)\;
		\((r_k,\Image_k)\gets\Execute{\(\Script_0\oplus\Delta_{\text{acc}}\)}\)\;
		\lIf{\(d(\Image_0,\Image_k)>\tau\)}{
			\Return \(\langle\Report,\Script_0\oplus\Delta_{\text{acc}}\rangle\)}
	}
	\Return \(\langle\Pass,\emptyset\rangle\)\;
\end{algorithm}

\section{Evaluation}
\label{sec:evaluation}

We evaluated \toolname via three research questions:

\begin{itemize}[topsep=0pt, leftmargin=*]
	\item \textbf{RQ1 (Bug Detection):} Can \toolname effectively find new bugs in real-world imperative \dataviz libraries?

	\item \textbf{RQ2 (Baseline Comparison):} How does \toolname compare against other testing techniques in terms of code coverage and bug detection?
%	\yq{I change the term ``the state of the art'' as I feel these are general tools.} \wq{ok.}

	\item \textbf{RQ3 (Ablation Study):} How much does each mutation operator contribute to \toolname's effectiveness?

\end{itemize}

\subsection{Experiment Setup}
\label{sec:setup}

\myparagraph{Hardware and software environment}
We conducted all experiments on a Linux server (\os, kernel \kernel) with a \cpu and \mem of memory, running \pyver. A \gpu was used only for the baseline experiments.

\myparagraph{Subjects}
We evaluated \toolname on three widely used Python \dataviz libraries:
\matplotlib \mplver, \bokeh \bokehver, and \plotly \plotlyver. \matplotlib
and \bokeh follow an imperative object model in which a script incrementally
mutates a hierarchy of graphic elements through sequences of API calls.
Although \plotly exposes a declarative specification interface, its internal
rendering updates can be modeled as imperative state transitions, so
\toolname applies to \plotly without modification. We excluded libraries that
are non-imperative or merely wrap another backend, such as \seaborn over
\matplotlib. All three subjects are actively maintained and still receive
recent bug reports, making them realistic bug-finding targets.
For each library, we collected seed scripts from
its official example gallery, because gallery examples are curated,
self-contained, and exercise diverse plot types and visual properties. After
discarding scripts that failed to execute, we collected \numseeds seeds:
\numseedmpl for \matplotlib, \numseedbokeh for \bokeh, and
\numseedplotly for \plotly. Each serves as a starting point \(\Script_0\) for
\toolname.
%We implemented \toolname in Python, realizing
%the full pipeline of \Cref{sec:approach}. The visual oracle uses the
%\texttt{ImageHash} library for the pHash computation.
We ran \toolname on each library with a budget of 120 hours.

\myparagraph{Parameter settings}
We fixed the per-seed mutation budget to $K=\budgetK$ rounds. A smaller $K$ raises script-loading cost, while a larger $K$ makes the reproducing script harder to minimize. Following the calibration in \Cref{subsec:oracle_cal}, we set the oracle threshold to \(\tau=2\), flagging any mutation with a pHash distance above \(\tau\) as a candidate visual defect.

\subsection{RQ1: Bug Detection}
\label{sec:rq1}

% TODO: The ``Rejected'' column requires explanation if kept (why were 15 reports rejected?).
% Suggestion: remove this column from the table and briefly discuss rejected reports in discussion.tex instead.

\begin{table}[t]
    \centering
    \caption{New bugs detected by \toolname across three libraries.
    	The column \textit{Pending Fix} denotes confirmed bugs for which developers have opened a pull request not yet merged.
%		\wq{todo: update numbers. remove rejected column?}
%		\yq{yeah. I think you can briefly mention a few common case that are rejected in discussion, without saying the actual number for now.}
%		\wq{Updated.}
		}
    \label{tab:rq1}
    \begin{tabular}{lcccc}
        \toprule
        Library & Reported & Confirmed & Pending Fix & Fixed \\
        \midrule
        \matplotlib & \newbugsmpl & \confirmedbugsmpl & \pendingbugsmpl & \fixedbugsmpl \\
        \bokeh      & \newbugsbokeh & \confirmedbugsbokeh & \pendingbugsbokeh & \fixedbugsbokeh \\
        \plotly     & \newbugsplotly & \confirmedbugsplotly & \pendingbugsplotly & \fixedbugsplotly \\
        \midrule
        Total       & \newbugs & \confirmedbugs & \pendingbugs & \fixedbugs \\
        \bottomrule
    \end{tabular}
\end{table}
\begin{figure*}[t]
	\centering
	\captionsetup[subfigure]{skip=2pt}
	% Each case: left = correct (expected), right = visual defect (actual).
	\begin{subfigure}{0.31\textwidth}
		\centering
		% trim=L B R T (pts; image is 460.8x345.6). Removes the left/top whitespace around the 3D box.
		\includegraphics[trim=50 20 80 50, clip, width=0.49\linewidth]{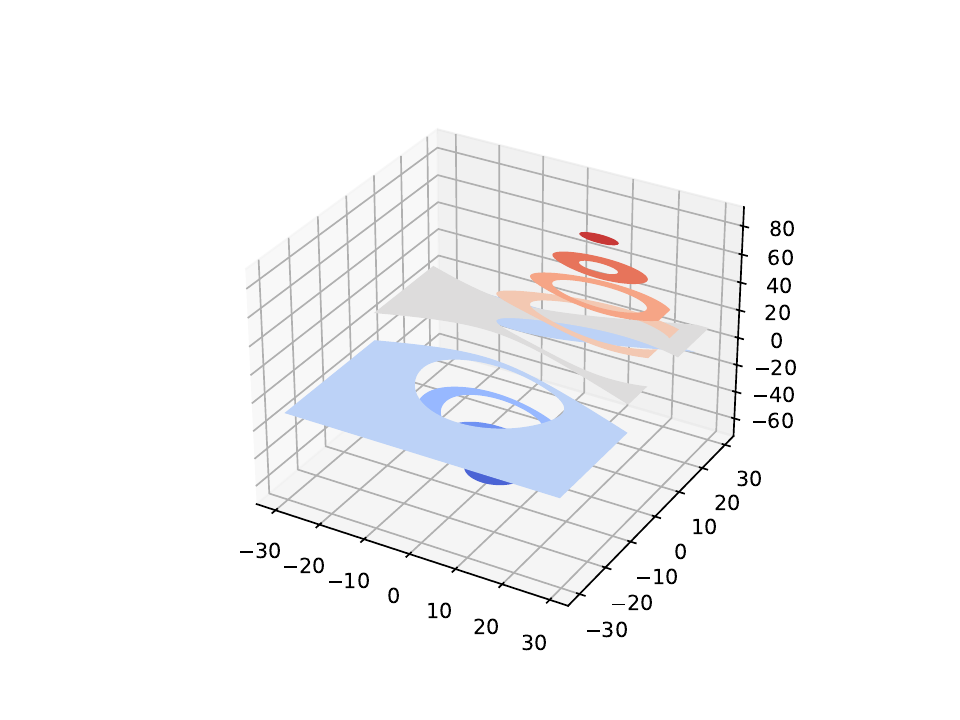}\hspace{2pt}%
		\includegraphics[trim=80 00 20 20, clip, width=0.49\linewidth]{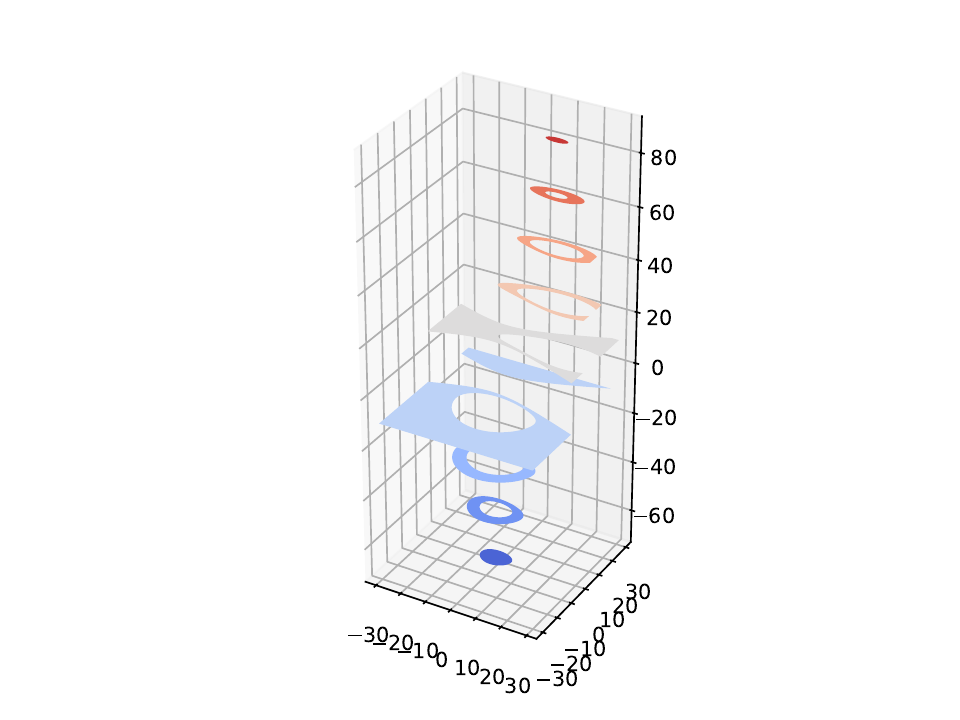}
		\subcaption{\matplotlib: aspect lost after set-revert.}
		\label{fig:case1}
	\end{subfigure}
	\hfill
	\begin{subfigure}{0.31\textwidth}
		\centering
		% trim=L B R T (pts; image is 350x350). Zooms into the bottom-left where x-axis labels overlap.
		\includegraphics[trim=0 0 150 150, clip, width=0.49\linewidth]{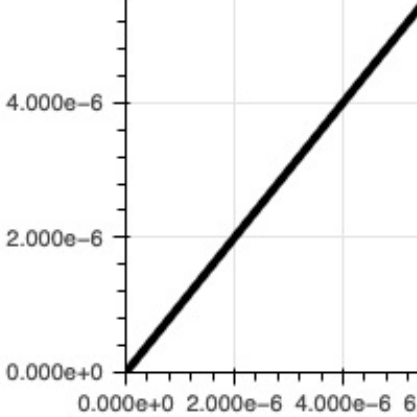}\hfill
		\includegraphics[trim=0 0 150 150, clip, width=0.49\linewidth]{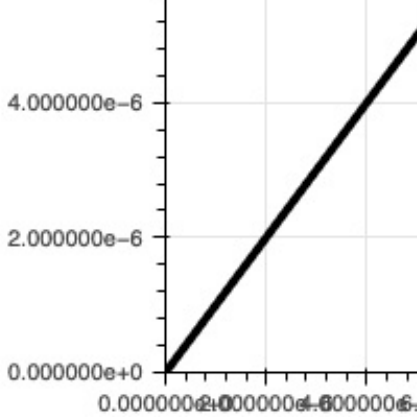}
		\subcaption{\bokeh: overlapping ticks after set-revert.}
		\label{fig:case2}
	\end{subfigure}
	\hfill
	\begin{subfigure}{0.36\textwidth}
		\centering
		% trim=L B R T (pts; image is 2544x1307). Near-square crop of the band-rich center-right region.
		\includegraphics[trim=1150 40 200 40, clip, width=0.45\linewidth]{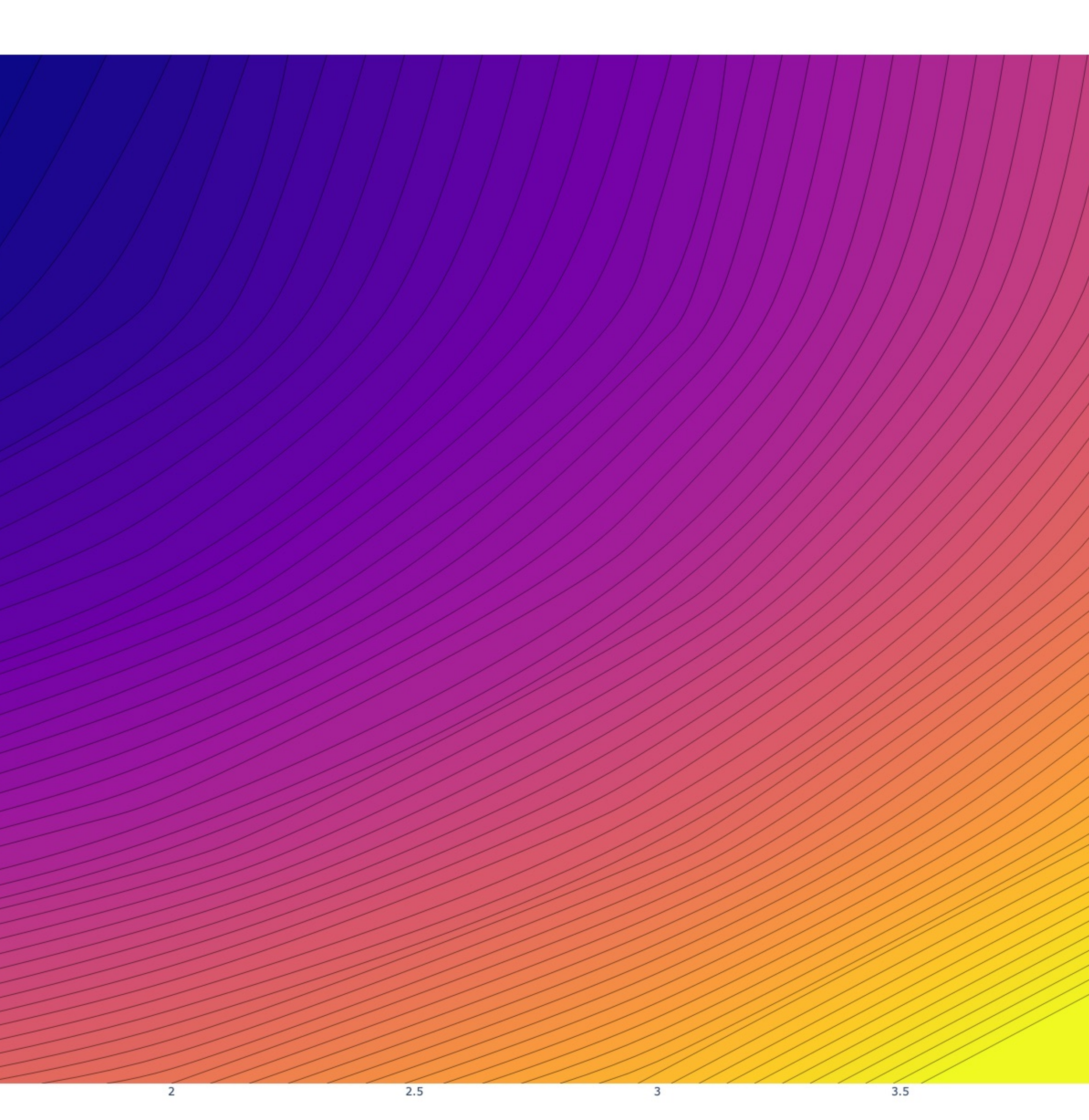}\hfill
		\includegraphics[trim=1150 40 200 40, clip, width=0.45\linewidth]{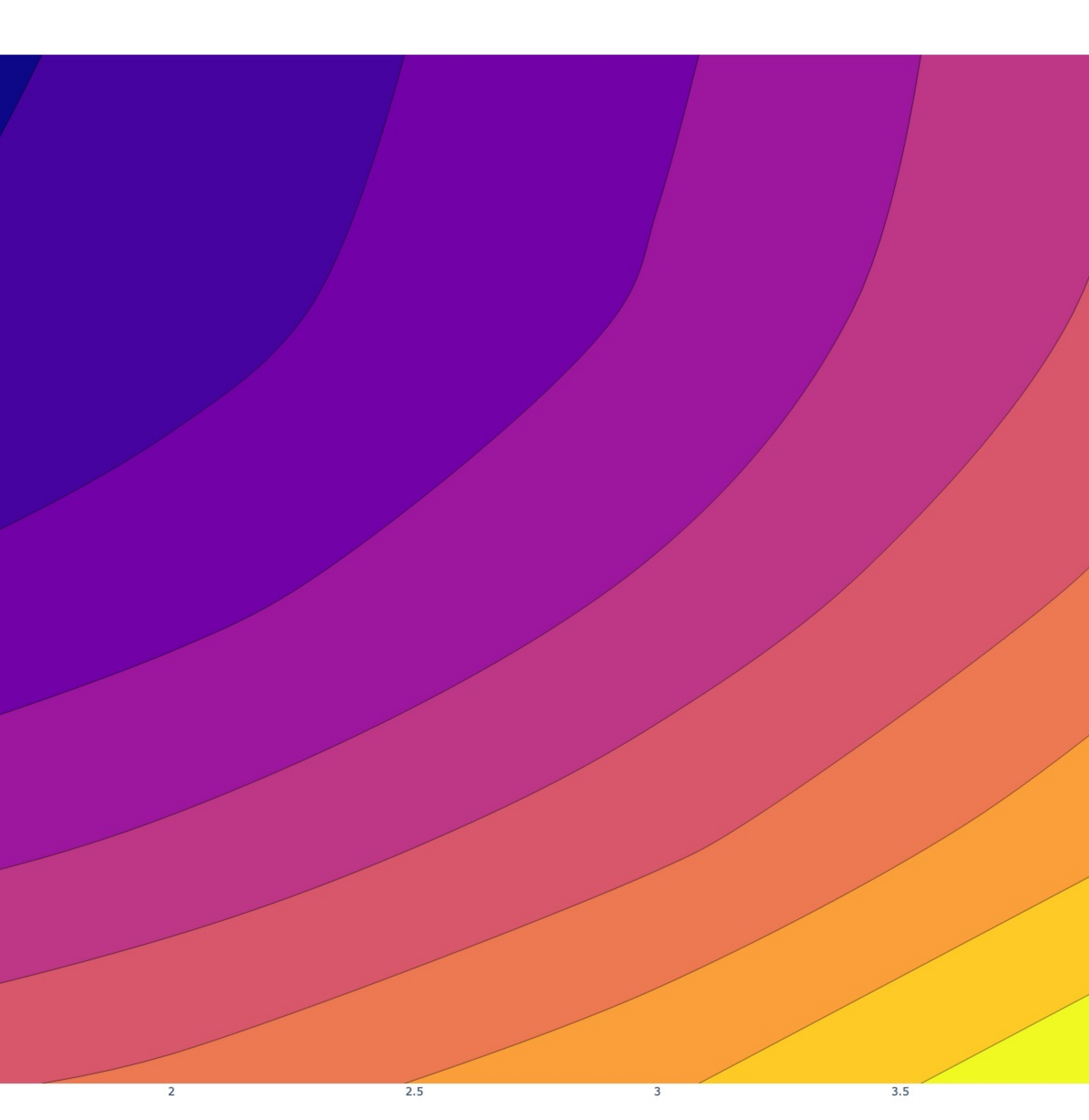}
		\subcaption{\plotly: contour size ignored in the seed render.}
		\label{fig:case3}
	\end{subfigure}

	\caption{Case studies of three confirmed bugs that \toolname detected across \matplotlib, \bokeh, and \plotly. In each pair, the left image is the correct render and the right image is the visual defect that the endpoint-preserving mutation reveals.}
	\label{fig:case-study}
\end{figure*}

%\myparagraph{Overall results}
We ran \toolname on the seed corpus of each library and manually triaged
every reproducing script emitted.
We report \newbugs previously unknown bugs to the GitHub repositories of the respective
libraries.
\Cref{tab:rq1} summarizes the outcome. Developers have confirmed
\confirmedbugs of these reports and already fixed \fixedbugs.
The rest are pending a fix or under triage. The confirmed bugs span all three libraries, showing that \toolname exposes bugs across diverse imperative \dataviz libraries.

\myparagraph{Symptoms}
The confirmed bugs fall into two symptom classes. \toolname found \confirmedbugsvisual
visual defects, where the library silently renders an incorrect plot, and
\confirmedbugscrash crashes, where a semantics-preserving mutation raises an exception.
Visual defects dominate. These silent corruptions produce no
error message and would escape any approach that detects crashes alone.
This confirms the value of comparing the rendered images of a script and
its endpoint-preserving mutant.

\myparagraph{Affected components}
The confirmed bugs span four categories of graphic
element. \toolname found \confirmedbugsencoding defects in visual encoding, \confirmedbugsannotaiton in annotations,
\confirmedbugslayout in layout, and \confirmedbugsscale in scales. This spread indicates that stateful
update bugs are not confined to a single subsystem but pervade
the rendering pipeline.

\myparagraph{Case studies}
We detail three confirmed bugs to illustrate how
endpoint-preserving mutations expose defects.
\Cref{fig:case-study} shows their correct and defective outputs.

\myparagraph{\matplotlib: 3D aspect ratio not restored}
On a filled 3D contour plot (Issue \#31276~\cite{githubBugSetting}, \cref{fig:case1}), a set-revert mutation sets
the axes aspect to \mycode{'equal'} and immediately reverts it to
\mycode{'auto'}. A correct library restores the projection, but
\matplotlib keeps the intermediate aspect and renders a distorted plot.
Developers confirmed the bug and opened a fix.

\myparagraph{\bokeh: scientific tick format not restored}
On a line drawn over a range on the order
of \(10^{-5}\), where the axis uses a compact scientific tick format
(Issue \#15031~\cite{githubCannotRecover}, \cref{fig:case2}), a set-revert mutation
toggles the \mycode{use_scientific} property off and then on. The
restored formatter falls back to a verbose format whose labels overlap,
which differs from the seed. The developers fixed and merged the bug.

\myparagraph{\plotly: contour size ignored in the seed}
On a contour plot whose contour size is
set to \mycode{0.25}, a value that the figure model registers correctly
(Issue \#5613~\cite{githubBUGContourssize}, \cref{fig:case3}), a redundant-set mutation
reassigns the same value. The seed render produces discrete bands, while the redundant-set render produces a smooth gradient (with finer contour lines). The mutation therefore exposes that the seed
render was already incorrect.
This case shows that an endpoint-preserving mutation validates not only the state transition it introduces, but also the initial state, since a buggy initial render becomes detectable once the mutation produces a divergent image.

\finding{\toolname discovers \newbugs previously unknown bugs across \matplotlib,
\bokeh, and \plotly, of which developers confirmed \confirmedbugs and already fixed \fixedbugs.
Most are silent visual defects that crash-based testing would miss.}

\begin{figure}[t]
	\centering
	\includegraphics[width=0.85\linewidth]{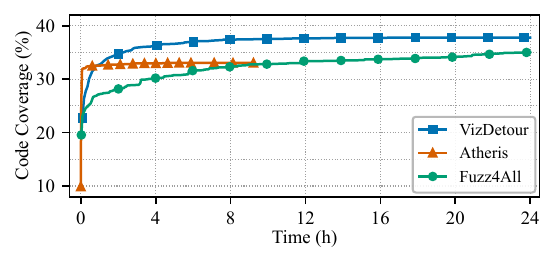}
	\caption{Line coverage on \matplotlib 3.9.2 over \timebudget hours. \baselineA terminates early due to memory exhaustion.}
	\label{fig:coverage}
\end{figure}

\subsection{RQ2: Baseline Comparison}
\label{sec:rq2}
%\yq{add a short opening explain why we select them (there is not existing baseline special targeted at xxxx so we selected two general fuzzing techniques xxxx)} \wq{added.}
As no existing fuzzing technique targets visual defects in \dataviz libraries, we adopted two general-purpose fuzzers as baselines. We compared \toolname against them on \matplotlib 3.9.2: \baselineA~\cite{atheris} (a byte-level coverage-guided fuzzer) and \baselineB~\cite{fuzz4all} (an LLM-based generation fuzzer). Both baselines can only flag scripts that raise runtime exceptions (\eg, recursion overflows, memory exhaustion, or deep exceptions in the rendering stack), as they rely on a crash oracle with no mechanism to detect silent visual defects. We initialized \baselineA with valid \matplotlib seed scripts and filtered shallow exceptions to focus on core rendering errors. For \baselineB, we used \texttt{starcoder2-7b}~\cite{starcoder2} and replayed all generated scripts against \matplotlib 3.11.0 to verify defects. Each tool got \timebudget hours on the same machine to measure coverage and historical bug detection.
%\yq{what oracle are used by each baseline?} \wq{added.}

\begin{figure}[t]
	\centering
	\includegraphics[width=.7\columnwidth]{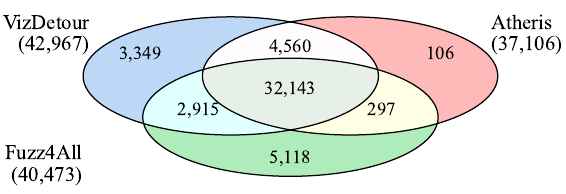}
	\caption{Coverage composition on \matplotlib 3.9.2. \toolname, \baselineA, and \baselineB cover 48{,}488 lines in total.}
	\label{fig:cov-composition}
\end{figure}

\myparagraph{Code Coverage}
\Cref{fig:coverage} shows that \toolname achieves the fastest growth and the highest final line coverage ($\sim$37\%). \baselineA quickly reaches $\sim$33\% due to its seed corpus but stagnates and crashes from memory exhaustion around hour nine, as byte-level mutations rarely produce valid programs. \baselineB grows steadily to $\sim$35\% but stays below \toolname.
\Cref{fig:cov-composition} compares the covered-line sets of the three tools. Although the tools share a common core of covered lines, their exclusive coverage differs significantly. \baselineB contributes the most exclusive coverage, but it concentrates in peripheral output backends (\eg, PDF and SVG exporters). In contrast, \toolname covers core rendering and layout modules (\eg, \texttt{patches}, \texttt{constrained\_layout}) where visual defects typically arise.
% Thus, while \baselineB covers more unique lines overall, it focuses on output generation rather than the stateful rendering and layout updates that \toolname primarily exercises.

\myparagraph{Historical Bug Detection}
%\Cref{tab:rq2-bugs} reports the bugs that each tool detects.
%Specifically,
We evaluate each tool against historical bugs present in \matplotlib 3.9.2 but fixed in the latest stable version 3.11.0.
\toolname detects three of these visual defects within the budget. \baselineA finds only one crash before running out of memory, and \baselineB finds none. This shows why a relative visual comparison matters: the baselines target crashes and miss the silent corruptions that dominate imperative \dataviz libraries.
%\yq{what do you mean by historical bug detection} \wq{clarified.}

\finding{\toolname achieves the highest and fastest line coverage, concentrated in core rendering modules. Aided by its visual oracle, \toolname detects three historical visual defects that neither baseline finds.}

%\yq{need a few examples of the real detected bug here, in addition to the motivating example}
%\wq{Added case studies of three confirmed bugs in \Cref{sec:rq1}.}
%\yq{I think we should add the link to bug report; or we at least can hid the bug number for now but say something like "the actual link to issue is hided for anonymous submission."} \wq{added.}

\subsection{RQ3: Ablation Study}
\label{sec:rq3}
\begin{table}[t]
	\centering
	\caption{Contribution of mutation operators in bug detection. The reduction relative to \toolname is in parentheses.}
	\label{tab:rq3-ablation}
	\renewcommand{\arraystretch}{0.85}
	\begin{tabular}{lcc}
		\toprule
		Variant & New bugs & Confirmed bugs \\
		\midrule
		\textbf{\toolname} & \textbf{47} & \textbf{39} \\
		w/o Set-Revert & 14 (-70.2\%) & 14 (-64.1\%) \\
		w/o Redundant-Set & 40 (-14.9\%) & 32 (-17.9\%) \\
		w/o Remove-Readd & 40 (-14.9\%) & 32 (-17.9\%) \\
		\bottomrule
	\end{tabular}
\end{table}
We evaluate the contribution of the three mutation operators by removing each component individually. \Cref{tab:rq3-ablation} reports the resulting drop in detected bugs.
%\yq{what do you mean by visual oracle here? If we remove it, what is used as oracle? According to claude, this term is never defined.}
%\wq{I defined visual oracle in \Cref{sec:oracle}. I removed the ablation study of visual oracle. Crash bug detection is not our focus, so visual oracle cannot be removed.}
%\yq{I see}
Among the mutation operators, removing \emph{set-revert} causes the sharpest decline in effectiveness. It reduces the new bugs by 70.2\% and the confirmed bugs by 64.1\%. Removing \emph{redundant-set} or \emph{remove-readd} each reduces the new bugs by 14.9\% and the confirmed bugs by 17.9\%. Though \emph{redundant-set} is a specialized case of \emph{set-revert}, we keep it separate because it is highly cost-effective. It simply reassigns the current value, eliminating the need to search for or generate alternative valid values. Each operator contributes uniquely to the overall effectiveness of \toolname, justifying its inclusion.
%\yq{actually we can report how many confirmed bugs are detected by each operator as well.} \wq{added.}

\finding{All three operators matter, but set-revert dominates: removing it cuts new bugs by 70.2\% and confirmed by 64.1\%, versus 14.9\% and 17.9\% for the other two.}

\section{Discussion}
\label{sec:discussion}

\subsection{Oracle Calibration}
\label{subsec:oracle_cal}

A naive threshold for the pHash distance (\eg, \(d>0\)) is impractical because floating-point non-determinism in rendering backends, sub-pixel anti-aliasing, and font rasterization produces non-zero distances even between semantically identical images. Instead, we calibrate the anomaly threshold \(\tau\) from the empirical distribution of pHash distances under mutations where no structural bug is present.

We executed our mutation pipeline across the collected seed scripts, evaluating 38,158 mutations to construct the empirical null distribution \(\NullDist\). Because the vast majority of mutations are strictly semantics-preserving, \(\NullDist\) is overwhelmingly concentrated at zero (38,096 cases, \({\sim}99.84\%\)). A minor, benign noise floor occurs at distance 2 (53 cases, \({\sim}0.14\%\)), representing imperceptible pixel rounding variances. Genuine structural anomalies emerge only in the sparse tail where \(d \ge 4\). To cleanly isolate these structural anomalies from rasterization noise, we set \(\tau = 2\), matching the 99th percentile boundary clear of the baseline noise floor. A mutant with \(d(\Image_0,\Image_k) > \tau\) is reported as suspicious.

\subsection{False Positives}
\label{subsec:false_positives}

Not every reproducing script \toolname reports witnesses a genuine defect. Developers rejected some reports as intended behavior, and these false positives follow three recurring patterns.
%\yq{my suggestion is not mention the concrete number of rejected bug. Just say a few examples are rejected as xxx.} \wq{updated.}
First, a mutation can combine two incompatible APIs. In Issue~\#31229~\cite{githubBugConstrainedPosition}, \mycode{set_position} disables constrained layout, so adjusting and restoring an axes position need not preserve the figure. Second, a getter and its setter can be asymmetric, so \mycode{set_x(get_x())} is not an identity. In Issue~\#31136~\cite{githubBugPolarSpine}, \mycode{get_transform} of a spine returns a composite transform whereas \mycode{set_transform} assigns only one component, recovered through \mycode{get_data_transform}. Third, a setter can carry a hidden side effect on coupled state. In Issue~\#31246~\cite{githubBugWrapReset}, enabling text wrapping also changes the rotation mode, so disabling wrapping alone does not restore the layout.

To filter these reports, \toolname currently relies on hardcoded blacklists of non-invertible API combinations. This static approach is brittle across library versions and misses deep, implicit side effects. A promising direction is to leverage LLM-powered semantic agents~\cite{he2025llmagent} that inspect the library's source code and documentation. By analyzing API implementations to deduce hidden state dependencies (\eg, recognizing that \mycode{set_wrap} implicitly modifies \mycode{rotation_mode}) and verifying whether a sequence is truly invertible per developer contracts, these agents can prune false positives and make \toolname a context-aware semantic oracle.

\subsection{Threats to Validity}
\label{sec:threats}

%\myparagraph{Internal validity}
The visual oracle is the primary internal threat: perceptual hashing can yield false positives from rasterization noise or false negatives from subtle deviations. We mitigate this by calibrating $\tau$ on \nullN null mutations (\Cref{subsec:oracle_cal}) and validating all defects with developers. Second, manual triage involves human judgment. We reduce this by minimizing mutation sequences to isolate root causes and submitting only distinct failures upstream. Finally, random element sampling may affect defect discovery, which we address by running each library for approximately 120 hours.

%\myparagraph{External validity}
The primary external threat is subject selection: we evaluate \toolname on three prominent Python libraries with imperative APIs. These cover the dominant imperative \dataviz paradigms in Python, but our findings may not generalize to declarative frameworks or other language ecosystems. Our seeds, harvested from official galleries, may not capture the complexity of real-world user scripts, but they provide an idiomatic baseline that maximizes API coverage while ensuring initial validity. Finally, because \toolname targets stateful update bugs via endpoint-preserving mutations, extending it to orthogonal defect categories, such as parameter validation errors or static rendering bugs, remains future work.

\section{Related Work}
\label{sec:related_work}

\myparagraph{Fuzzing}
Fuzzing automatically generates test inputs that expose software defects~\cite{DBLP:journals/software/BohmeCR21}. Mutation-based fuzzers perturb existing inputs, while generation-based fuzzers synthesize inputs from a grammar or model. Coverage-guided engines such as \baselineA~\cite{atheris} mutate inputs under code-coverage feedback. Compiler testing has relied on these strategies to generate random programs~\cite{csmith} and catch bugs via differential or metamorphic oracles~\cite{emi,ccmd}. Recent work applies language models to fuzzing, including \baselineB for universal input generation~\cite{fuzz4all}, CovRL-Fuzz for JS interpreters~\cite{eom2024fuzzing}, and COMFUZZ for compilers~\cite{ye2023comfuzz}. Most of these fuzzers target crashes or differential mismatches across independent implementations. In contrast, \toolname requires no second implementation, targeting a single \dataviz library by deriving a relative oracle via endpoint-preserving mutations.

\myparagraph{GUI and rendering testing}
GUI testing explores application states via simulated interactions to find functional and visual faults~\cite{chang2010guicv, su2017stochasticguitesting, wang2025llmdroid}. In contrast, \toolname mutates plotting API calls rather than interactive event sequences. Closest to our work, Janus detects rendering bugs across browsers~\cite{zhou2025janus}, while Metamong targets inconsistencies between initial browser rendering and render-update behavior~\cite{song2023metamong}. Like them, \toolname assumes equivalent inputs yield consistent visuals. However, browsers are governed by shared standards, whereas \dataviz libraries have no common specifications, ruling out differential testing. \toolname therefore derives a relative oracle from the library's own stateful semantics, enabling testing of a single implementation.

\section{Conclusion}
\label{sec:conclusion}

We presented \toolname, an automated testing approach for detecting silent
visual defects in imperative \dataviz libraries. Using endpoint-preserving
mutation, \toolname compares a seed script with a semantics-equivalent mutant,
turning the missing-oracle problem into a relative visual comparison.
Applied to \matplotlib, \bokeh, and \plotly, \toolname discovers \newbugs previously
unknown bugs, with \confirmedbugs confirmed by developers and \fixedbugs already fixed. It also
achieves higher coverage than established, domain-agnostic fuzzers and exposes visual
defects they miss.
These results show that endpoint-preserving mutation tests
visualization libraries without an absolute oracle. Future
work will extend it to declarative libraries, additional language
ecosystems, and richer semantic oracles and triage support.

%\section{Data Availability}
%\label{sec:data_availability}
%We make our artifact, including the experiment results and the replication code, publicly available at~\url{https://github.com/smith2936/vizdetour} to facilitate follow-up research studies.

%\input{acknowledgment}

\bibliographystyle{IEEEtran}
\bibliography{references}

% \vspace{12pt}
% \color{red}
% IEEE conference templates contain guidance text for composing and formatting conference papers. Please ensure that all template text is removed from your conference paper prior to submission to the conference. Failure to remove the template text from your paper may result in your paper not being published.

\end{document}